\documentclass{article}
\usepackage{arxiv}
\usepackage[utf8]{inputenc}
\usepackage[T1]{fontenc}
\usepackage{hyperref}
\usepackage{url}
\usepackage{booktabs}
\usepackage{amsfonts}
\usepackage{amsmath}
\usepackage{amssymb}
\usepackage{nicefrac}
\usepackage{microtype}
\usepackage{cleveref}
\usepackage{graphicx}
\usepackage{natbib}
\usepackage{doi}
\usepackage{bm}
\usepackage{multirow}
\usepackage{subcaption}

\graphicspath{{figures/}}

\title{Nonlinear Model Order Reduction for Coupled Aeroelastic-Flight Dynamic Systems}

\author{
  Nikolaos D.~Tantaroudas\thanks{Corresponding author. Senior Researcher, ICCS.} \\
  Institute of Communications and Computer Systems (ICCS)\\
  9 Iroon Politechniou Street, Zografou, Athens 15773, Greece \\
  \texttt{nikolaos.tantaroudas@iccs.gr} \\
  \And
  Ilias Karachalios \\
  National Technical University of Athens\\
  Zografou, 157 80, Athens, Greece \\
}

\renewcommand{\shorttitle}{Nonlinear Model Order Reduction for Aeroelastic Systems}

\hypersetup{
  pdftitle={Nonlinear Model Order Reduction for Coupled Aeroelastic-Flight Dynamic Systems},
  pdfsubject={physics.flu-dyn, cs.CE},
  pdfauthor={N.D. Tantaroudas, I. Karachalios},
  pdfkeywords={Model order reduction, Aeroelasticity, Flexible aircraft, Nonlinear dynamics, Taylor series expansion}
}

\begin{document}
\maketitle

\begin{abstract}
A systematic approach to nonlinear model order reduction (NMOR) of coupled fluid-structure-flight dynamics systems of arbitrary fidelity is presented. The technique employs a Taylor series expansion of the nonlinear residual around equilibrium states, retaining up to third-order terms, and projects the high-dimensional system onto a small basis of eigenvectors of the coupled-system Jacobian matrix. The biorthonormality of right and left eigenvectors ensures optimal projection, while higher-order operators are computed via matrix-free finite difference approximations. The methodology is validated on three test cases of increasing complexity: a three-degree-of-freedom aerofoil with nonlinear stiffness (14 states reduced to 4), a HALE aircraft configuration (2,016 states reduced to 9), and a very flexible flying-wing (1,616 states reduced to 9). The reduced-order models achieve computational speedups of up to 600 times while accurately capturing the nonlinear dynamics, including large wing deformations exceeding 10\% of the wingspan. The second-order Taylor expansion is shown to be sufficient for describing cubic structural nonlinearities, eliminating the need for third-order terms. The framework is independent of the full-order model formulation and applicable to higher-fidelity aerodynamic models.
\end{abstract}


\section*{Nomenclature}

\subsection*{Abbreviations}
\begin{tabular}{@{}ll@{}}
DOF & Degree of freedom \\
FE & Finite element \\
FOM & Full-order model \\
HALE & High-altitude long-endurance \\
HTP & Horizontal tail plane \\
LCO & Limit-cycle oscillation \\
MOR & Model order reduction \\
NFOM & Nonlinear full-order model \\
NMOR & Nonlinear model order reduction \\
NROM & Nonlinear reduced-order model \\
POD & Proper orthogonal decomposition \\
ROM & Reduced-order model \\
UAV & Unmanned aerial vehicle \\
UVLM & Unsteady vortex-lattice method \\
VFA & Very flexible aircraft \\
VTP & Vertical tail plane \\
\end{tabular}

\subsection*{Roman Symbols}
\begin{tabular}{@{}ll@{}}
$a$ & Non-dimensional elastic axis position (from mid-chord) \\
$b$ & Semi-chord length \\
$\mathbf{A}$ & Jacobian matrix of the coupled system \\
$\mathbf{B}_c, \mathbf{B}_g$ & Control and gust input matrices \\
$C(k)$ & Theodorsen function \\
$C_L, C_m, C_h$ & Lift, moment, and hinge moment coefficients \\
$D_{kij}$ & Second-order interaction coefficients \\
$E_{kijl}$ & Third-order interaction coefficients \\
$EI_2$ & Flapwise bending stiffness \\
$EI_3$ & Chordwise bending stiffness \\
$GJ$ & Torsional stiffness \\
$k$ & Reduced frequency, $\omega b / U$ \\
$m$ & Number of retained modes in ROM \\
$n$ & Dimension of full-order state vector \\
$\mathbf{R}$ & Nonlinear residual vector \\
$U$ & Freestream velocity \\
$U^*$ & Non-dimensional freestream velocity \\
$\mathbf{u}_c$ & Control input vector \\
$\mathbf{u}_d$ & Gust disturbance input vector \\
$\mathbf{w}$ & State vector \\
$\mathbf{w}_0$ & Equilibrium (trim) state \\
$\mathbf{z}$ & Reduced state vector \\
\end{tabular}

\subsection*{Greek Symbols}
\begin{tabular}{@{}ll@{}}
$\alpha$ & Angle of attack / pitch degree of freedom \\
$\gamma$ & $\mathcal{H}_\infty$ performance bound \\
$\boldsymbol{\gamma}$ & Beam axial strain vector \\
$\delta$ & Trailing-edge flap deflection \\
$\epsilon$ & Finite-difference perturbation parameter \\
$\boldsymbol{\zeta}$ & Quaternion attitude vector \\
$\boldsymbol{\kappa}$ & Beam curvature vector \\
$\lambda_i$ & Eigenvalue of the Jacobian \\
$\xi$ & Plunge degree of freedom \\
$\rho$ & Air density \\
$\sigma, \sigma_1, \sigma_2$ & Flexibility (stiffness scaling) parameters \\
$\tau$ & Non-dimensional aerodynamic time \\
$\boldsymbol{\phi}_i$ & Right eigenvector \\
$\boldsymbol{\psi}_i$ & Left eigenvector \\
$\boldsymbol{\Phi}$ & Matrix of right eigenvectors \\
$\boldsymbol{\Psi}$ & Matrix of left eigenvectors \\
$\phi_w$ & Wagner indicial function \\
$\psi_k$ & K\"ussner indicial function \\
$\boldsymbol{\omega}_B$ & Angular velocity vector in body frame \\
\end{tabular}

\subsection*{Operators}
\begin{tabular}{@{}ll@{}}
$\mathcal{B}(\cdot, \cdot)$ & Symmetric bilinear operator (second-order) \\
$\mathcal{C}(\cdot, \cdot, \cdot)$ & Symmetric trilinear operator (third-order) \\
\end{tabular}

\section{Introduction}
\label{sec:intro}

The design of next-generation high-altitude long-endurance (HALE) and solar-powered aircraft poses fundamental challenges to established aeroelastic and flight dynamic analysis methods. These vehicles feature very high-aspect-ratio wings that undergo large structural deformations during flight, creating strong coupling between structural flexibility, unsteady aerodynamics, and rigid-body flight dynamics~\citep{Patil2001, Patil2006, Noll2004}. The 2003 NASA Helios prototype accident, in which a HALE platform with 75-m wingspan experienced catastrophic failure following an encounter with atmospheric turbulence, demonstrated that the traditional separation between flight dynamics and aeroelastic analysis is inappropriate for such configurations~\citep{Noll2004}. The mishap investigation revealed that the vehicle's large wing deformations, exceeding 40\,ft at the tip prior to the instability, produced a fundamentally different dynamic response than predicted by uncoupled models.

Several research groups have since developed coupled aeroelastic-flight dynamic toolboxes for very flexible aircraft. \citet{Patil2001} and \citet{Patil2006} pioneered the investigation of flight dynamics for Helios-like flying wings using geometrically-exact beam theory coupled with two-dimensional finite-state aerodynamics. Their work showed that wing flexibility modifies phugoid and short-period modes, potentially leading to instabilities not predicted by rigid-body analysis. \citet{Murua2012} coupled an unsteady vortex-lattice method (UVLM) with a geometrically-exact displacement-based beam, providing higher-fidelity aerodynamic modelling. \citet{Hesse2014} developed reduced-order aeroelastic models for dynamics of manoeuvring flexible aircraft using a mean-axes approach. \citet{SuCesnik2010} examined nonlinear aeroelasticity and flight dynamics of high-altitude long-endurance aircraft, highlighting the importance of geometric stiffening effects at large deformations. These full-order models (FOMs), while capturing essential physics, result in systems with hundreds to thousands of degrees of freedom (DOF), making parametric studies and control design computationally prohibitive. A single nonlinear time-domain simulation can require several hours of wall-clock time, and certification-relevant parameter sweeps demand hundreds of such simulations.

Model order reduction (MOR) offers a path to overcome this computational barrier. Various approaches have been explored, including proper orthogonal decomposition (POD)~\citep{Lucia2004}, harmonic balance methods~\citep{Badcock2011}, balanced truncation, and time-domain identification techniques~\citep{Dowell2001}. While effective in specific contexts, POD requires extensive training data that depend on the particular excitation; harmonic balance is restricted to periodic or quasi-periodic regimes; and linear identification cannot capture essential nonlinearities such as geometric stiffening and limit-cycle oscillations.

More recently, significant progress has been made in model reduction for geometrically nonlinear aeroelastic systems. \citet{Riso2023joa} systematically assessed the impact of low-order modelling on aeroelastic predictions for very flexible wings, demonstrating that a geometrically nonlinear beam coupled with potential flow strip theory can capture flutter onset and post-flutter behaviour with accuracy comparable to higher-fidelity models based on the unsteady vortex-lattice method. The same authors~\citep{Riso2023jfs} further investigated geometrically nonlinear effects in wing aeroelastic dynamics at large deflections, quantifying the conditions under which linear and nonlinear structural models yield divergent predictions. \citet{Goizueta2022} developed adaptive sampling strategies for interpolation of parametric reduced-order aeroelastic models, enabling efficient exploration of large design spaces and multiple flight conditions via Krylov-subspace projection. \citet{Candon2024} proposed the parameterisation of nonlinear aeroelastic reduced-order models through direct interpolation of Taylor partial derivatives, demonstrating that sparse higher-order polynomial representations achieve excellent precision in modelling high-amplitude limit-cycle oscillations while significantly reducing the required training data.

The present work employs a Taylor series-based nonlinear model order reduction (NMOR) approach that projects the coupled system onto eigenvectors of the Jacobian matrix~\citep{DaRonch2013control, DaRonch2013gust, DaRonch2012rom}. This approach retains nonlinear terms systematically through second- and third-order Taylor expansion coefficients, produces models that are excitation-independent once generated, and enables rapid parametric studies on a compact system. The methodology was first applied to gust loads analysis in~\citep{DaRonch2013gust, DaRonch2013control}, and was subsequently extended to free-flying flexible aircraft with coupled flight dynamics in~\citep{Tantaroudas2015scitech, Tantaroudas2014aviation}. Experimental validation against wind tunnel testing was performed in~\citep{DaRonch2014flutter, Papatheou2013ifasd, Fichera2014isma}, and a comprehensive overview of the framework is provided in the book chapter~\citep{Tantaroudas2017bookchapter}.

The contributions of this paper are: (i) a detailed presentation and validation of the NMOR framework across three test cases of increasing complexity, demonstrating reductions from $\sim$1,600 DOF to 9 with up to 600$\times$ computational speedup; (ii) a systematic demonstration that the second-order Taylor expansion is sufficient for cubic structural nonlinearities; (iii) a rigorous eigenvector basis selection strategy combining rigid-body and structural modes; and (iv) evidence that the methodology is independent of the full-order model formulation, applicable to any system expressible in first-order state-space form.

\section{Full-Order Model Formulation}
\label{sec:fom}

\begin{figure}[htbp]
\centering
\includegraphics[width=0.65\textwidth]{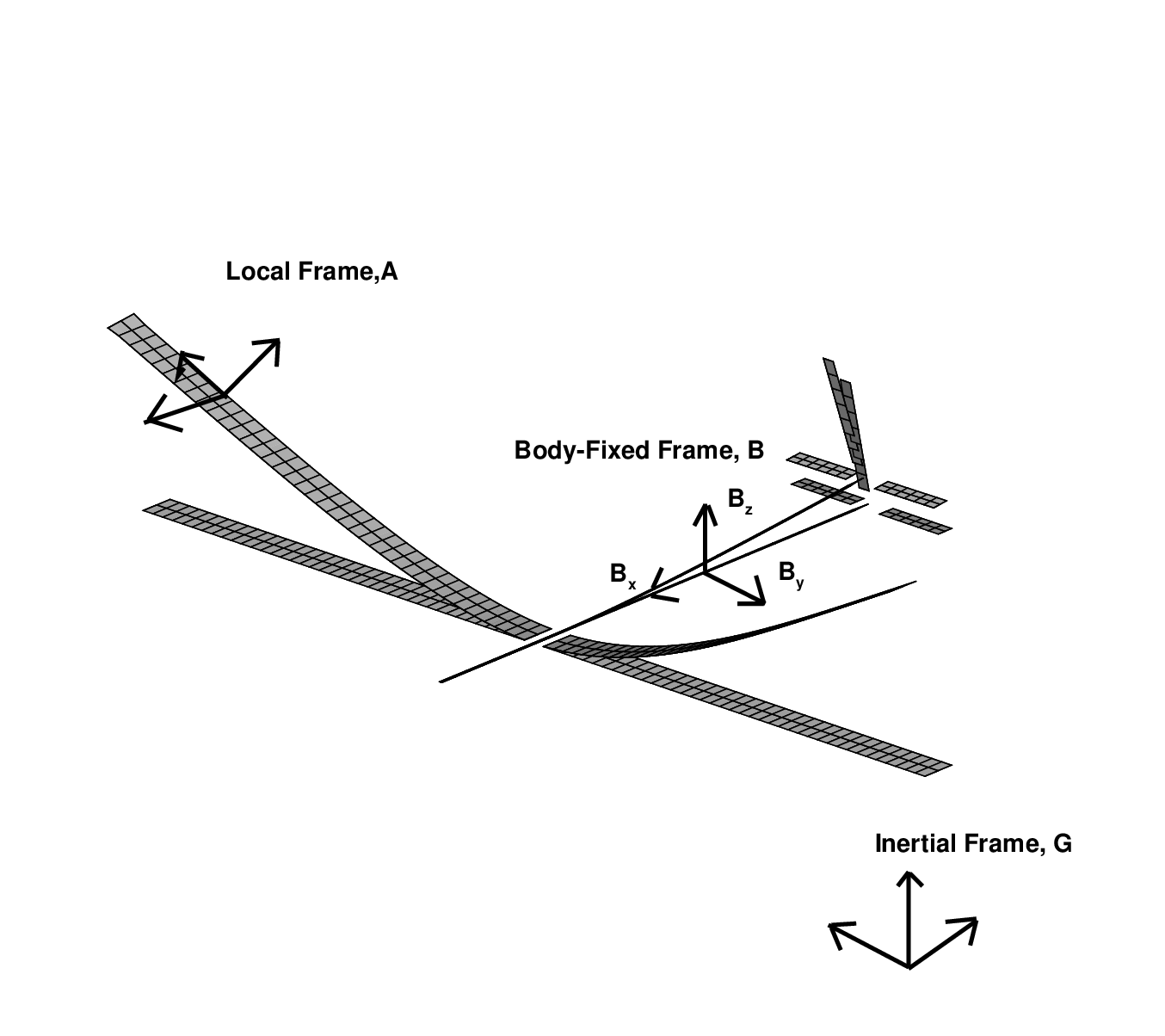}
\caption{Reference frames for the coupled aeroelastic-flight dynamic model: global inertial frame ($G$), body-fixed frame ($B$), and local aerodynamic frames ($A_i$) at each strip along the wing span. The structural beam axis deforms within the body frame, while aerodynamic loads are resolved in local frames aligned with the deformed beam cross-section.}
\label{fig:flight_frame}
\end{figure}

The coupled aeroelastic-flight dynamic system integrates three subsystems: unsteady aerodynamics, nonlinear structural dynamics, and rigid-body flight dynamics. The kinematic framework relating these subsystems is illustrated in~\Cref{fig:flight_frame}, which shows the global, body, and aerodynamic reference frames and the coordinate transformations between them.

\subsection{General Framework}

The coupled system is expressed as a set of nonlinear ordinary differential equations in first-order state-space form:
\begin{equation}
\frac{d\mathbf{w}}{dt} = \mathbf{R}(\mathbf{w}, \mathbf{u}_c, \mathbf{u}_d)
\label{eq:state_space}
\end{equation}
where $\mathbf{w} \in \mathbb{R}^n$ is the state vector of dimension $n$, $\mathbf{R}: \mathbb{R}^n \to \mathbb{R}^n$ is the nonlinear residual vector encoding the coupled dynamics, $\mathbf{u}_c \in \mathbb{R}^{n_c}$ is the control input vector (e.g., trailing-edge flap deflections), and $\mathbf{u}_d \in \mathbb{R}^{n_d}$ is the external disturbance vector (e.g., atmospheric gust velocity). The state vector is partitioned as:
\begin{equation}
\mathbf{w} = \left\{ \mathbf{w}_f, \; \mathbf{w}_s, \; \mathbf{w}_r \right\}^T
\end{equation}
where $\mathbf{w}_f \in \mathbb{R}^{n_f}$ denotes augmented aerodynamic states arising from indicial function approximations, $\mathbf{w}_s \in \mathbb{R}^{n_s}$ the structural degrees of freedom (nodal displacements and velocities), and $\mathbf{w}_r \in \mathbb{R}^{n_r}$ the rigid-body states encompassing translational velocity, angular velocity, position, and attitude quaternion.

At an equilibrium (or trim) state $\mathbf{w}_0$, the residual vanishes:
\begin{equation}
\mathbf{R}(\mathbf{w}_0, \mathbf{u}_{c0}, \mathbf{u}_{d0}) = \mathbf{0}
\end{equation}
The equilibrium is determined by solving the steady-flight trim problem, which in general accounts for the static aeroelastic deformation under flight loads. For very flexible aircraft, the trim deformation can exceed 10--25\% of the wingspan, requiring a fully nonlinear trim solution~\citep{Patil2006, Tantaroudas2017bookchapter}.

\subsection{Unsteady Aerodynamic Model}
\label{sec:aero}

\begin{figure}[htbp]
\centering
\includegraphics[width=0.55\textwidth]{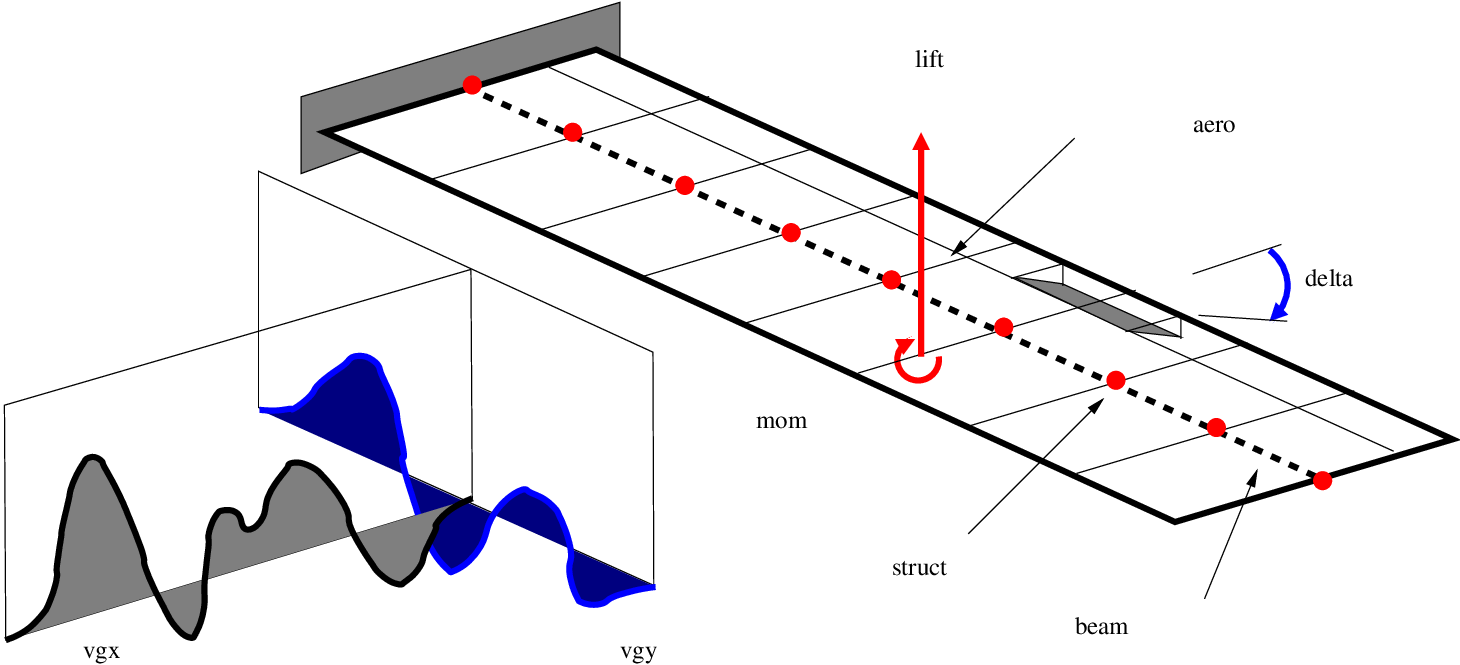}
\caption{Wing cross-section schematic showing the elastic axis location, trailing-edge flap, and aerodynamic strip definition. The section motion (pitch $\alpha$, plunge $\xi$), flap deflection $\delta$, and gust velocity $w_g$ each contribute independently to the total aerodynamic loading.}
\label{fig:wing_layout}
\end{figure}

The aerodynamic loads are computed using two-dimensional unsteady strip theory based on Theodorsen's thin-aerofoil framework~\citep{Theodorsen1935}. Each spanwise section is treated independently, with loads resolved in local aerodynamic frames attached to the deformed beam cross-section as shown in~\Cref{fig:wing_layout}. The total aerodynamic force and moment coefficients at each section decompose into contributions from three physically distinct mechanisms~\citep{DaRonch2014scitech_flight}:
\begin{equation}
C_i = C_{i,\text{motion}} + C_{i,\text{flap}} + C_{i,\text{gust}}, \qquad i = L, m, h
\label{eq:aero_decomp}
\end{equation}
where $L$, $m$, and $h$ denote lift, pitching moment, and hinge moment, respectively.

\subsubsection{Section motion contribution}

The circulatory lift due to section motion is expressed in terms of the effective angle of attack through Theodorsen's function $C(k)$, where $k = \omega b / U$ is the reduced frequency:
\begin{equation}
C_{L,\text{motion}} = 2\pi \left[ C(k)\left(\alpha + \frac{\dot{\xi}}{U} + \left(\frac{1}{2} - a\right)\frac{b\dot{\alpha}}{U}\right) + \frac{b\dot{\alpha}}{2U} \right] + \pi \frac{b}{U}\left(\dot{\alpha} + \frac{\ddot{\xi}}{U} - a\frac{b\ddot{\alpha}}{U}\right)
\label{eq:CL_motion}
\end{equation}
where $a$ is the non-dimensional elastic axis position (measured from mid-chord, positive aft), $b$ is the semi-chord, and $U$ is the freestream velocity.

For time-domain simulation, the frequency-domain Theodorsen function is replaced by the Wagner indicial response function~\citep{Wagner1925, Jones1938}:
\begin{equation}
\phi_w(\tau) = 1 - \Psi_1 e^{-\varepsilon_1 \tau} - \Psi_2 e^{-\varepsilon_2 \tau}
\label{eq:wagner}
\end{equation}
with coefficients $\Psi_1 = 0.165$, $\Psi_2 = 0.335$, $\varepsilon_1 = 0.0455$, and $\varepsilon_2 = 0.3$, where $\tau = Ut/b$ is the non-dimensional aerodynamic time. This two-exponential approximation~\citep{Jones1938} introduces two augmented aerodynamic states per strip for section motion effects, governed by:
\begin{equation}
\dot{x}_{w,j}(\tau) = -\varepsilon_j \frac{U}{b} x_{w,j}(\tau) + \frac{3}{4}\text{-chord downwash}, \qquad j = 1,2
\label{eq:aero_states_wagner}
\end{equation}

\subsubsection{Atmospheric gust contribution}

Atmospheric gust penetration is modelled through the K\"ussner indicial function~\citep{Kussner1936}, which describes the unsteady lift build-up on a thin aerofoil entering a sharp-edged gust:
\begin{equation}
\psi_k(\tau) = 1 - \Psi_3 e^{-\varepsilon_3 \tau} - \Psi_4 e^{-\varepsilon_4 \tau}
\label{eq:kussner}
\end{equation}
with $\Psi_3 = 0.5792$, $\Psi_4 = 0.4208$, $\varepsilon_3 = 0.1393$, and $\varepsilon_4 = 1.802$. This approximation adds two further augmented aerodynamic states per strip for gust coupling. The gust-induced lift coefficient at each strip is:
\begin{equation}
C_{L,\text{gust}}(\tau) = 2\pi \int_0^{\tau} \frac{d\psi_k}{d\sigma} \frac{w_g(\tau - \sigma)}{U} d\sigma
\end{equation}
where $w_g$ is the vertical gust velocity at the strip location.

Each spanwise strip therefore contributes 4 augmented aerodynamic states (2 from Wagner, 2 from K\"ussner) to the global state vector, plus additional states if flap aerodynamics are included. For a model with $N_s$ strips, the aerodynamic partition has dimension $n_f = 4 N_s$ (without flap) or $n_f = 8 N_s$ (with trailing-edge flap, which introduces its own indicial states).

\subsection{Structural Dynamics}
\label{sec:structure}

\begin{figure}[htbp]
\centering
\includegraphics[width=0.55\textwidth]{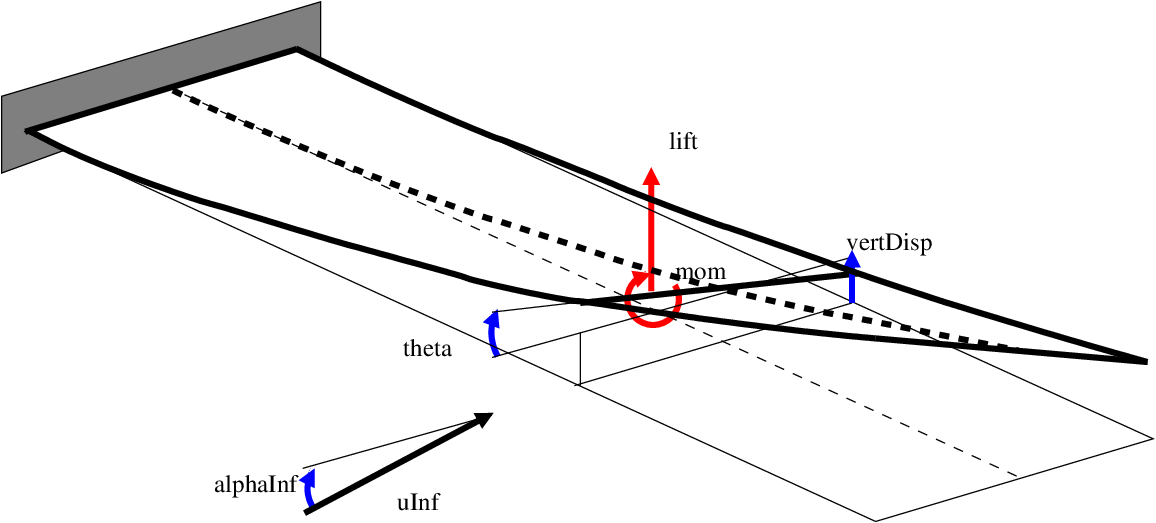}
\caption{Schematic of wing structural deformation showing the undeformed beam reference line and the deformed configuration. The local cross-sectional frame at each beam node rotates and translates with the deformation, producing curvature and twist that enter the strain-displacement relations of the geometrically-exact beam theory.}
\label{fig:wing_deformation}
\end{figure}

The structural dynamics are governed by geometrically-exact nonlinear beam equations~\citep{Hodges2003, Palacios2010} discretised using two-noded displacement-based finite elements, each node carrying six degrees of freedom, three translational and three rotational. The wing deformation schematic is shown in~\Cref{fig:wing_deformation}. The displacement field of each node $i$ relative to the undeformed configuration is:
\begin{equation}
\mathbf{q}_i = \left\{ p_{x,i}, \; p_{y,i}, \; p_{z,i}, \; \theta_{x,i}, \; \theta_{y,i}, \; \theta_{z,i} \right\}^T
\end{equation}
The rotations are parameterised by the Cartesian rotation vector, from which the rotation tensor $\mathbf{T}(\boldsymbol{\theta}_i)$ is obtained via the Rodrigues formula.

The beam strain measures, axial strain $\boldsymbol{\gamma}$ and curvature $\boldsymbol{\kappa}$, are computed from the nodal displacements using:
\begin{align}
\boldsymbol{\gamma} &= \mathbf{T}^T \mathbf{r}' - \mathbf{e}_1 \label{eq:strain_gamma}\\
\boldsymbol{\kappa} &= \mathbf{T}^T \mathbf{T}' \label{eq:strain_kappa}
\end{align}
where $\mathbf{r}' = \partial \mathbf{r}/\partial s$ is the derivative of the position vector along the beam reference line, $\mathbf{e}_1 = \{1, 0, 0\}^T$, and the prime denotes differentiation with respect to the curvilinear coordinate $s$. These strain measures are frame-invariant and account for arbitrary large displacements and rotations.

The virtual work principle yields the finite element equations of motion:
\begin{equation}
\mathbf{M}(\mathbf{w}_s) \begin{Bmatrix} \ddot{\mathbf{w}}_s \\ \ddot{\mathbf{w}}_r \end{Bmatrix} + \mathbf{Q}_{\text{gyr}} \begin{Bmatrix} \dot{\mathbf{w}}_s \\ \dot{\mathbf{w}}_r \end{Bmatrix} + \mathbf{Q}_{\text{stiff}} \begin{Bmatrix} \mathbf{w}_s \\ \mathbf{w}_r \end{Bmatrix} = \mathbf{R}_F
\label{eq:structural}
\end{equation}
where $\mathbf{M}$ is the tangent mass matrix, which depends on the structural states for the geometrically nonlinear formulation due to the configuration-dependent inertia terms; $\mathbf{Q}_{\text{gyr}}$ contains gyroscopic and Coriolis contributions arising from the coupling between structural and rigid-body velocities; $\mathbf{Q}_{\text{stiff}}$ contains the elastic internal forces and geometric stiffening terms; and $\mathbf{R}_F$ is the vector of generalised aerodynamic forces obtained from the strip theory model of~\Cref{sec:aero}.

\subsection{Flight Dynamics}

The rigid-body flight dynamics are described by Newton--Euler equations expressed in the body-fixed frame~$B$:
\begin{align}
m \left(\dot{\mathbf{v}}_B + \boldsymbol{\omega}_B \times \mathbf{v}_B\right) &= \mathbf{F}_{\text{aero}} + \mathbf{F}_{\text{grav}} + \mathbf{F}_{\text{thrust}} \label{eq:translational}\\
\mathbf{I}_B \dot{\boldsymbol{\omega}}_B + \boldsymbol{\omega}_B \times \mathbf{I}_B \boldsymbol{\omega}_B &= \mathbf{M}_{\text{aero}} + \mathbf{M}_{\text{grav}} \label{eq:rotational}
\end{align}
where $m$ is the total mass, $\mathbf{v}_B$ and $\boldsymbol{\omega}_B$ are the translational and angular velocity vectors in the body frame, $\mathbf{I}_B$ is the inertia tensor, and the right-hand sides collect the aerodynamic, gravitational, and propulsive generalised forces.

Attitude propagation employs quaternions $\boldsymbol{\zeta} = \{\zeta_0, \zeta_1, \zeta_2, \zeta_3\}^T$ to avoid the singularities inherent in Euler-angle representations~\citep{Tantaroudas2015scitech}:
\begin{equation}
\dot{\boldsymbol{\zeta}} = \frac{1}{2} \boldsymbol{\Omega}(\boldsymbol{\omega}_B) \boldsymbol{\zeta}, \qquad
\boldsymbol{\Omega}(\boldsymbol{\omega}) = \begin{bmatrix} 0 & -\omega_x & -\omega_y & -\omega_z \\ \omega_x & 0 & \omega_z & -\omega_y \\ \omega_y & -\omega_z & 0 & \omega_x \\ \omega_z & \omega_y & -\omega_x & 0 \end{bmatrix}
\label{eq:quaternion}
\end{equation}
with the unit-norm constraint $|\boldsymbol{\zeta}| = 1$ enforced at each time step.

\subsection{Coupled System in First-Order Form}

Combining the aerodynamic state equations~\eqref{eq:aero_states_wagner}, the structural equations~\eqref{eq:structural}, the flight dynamic equations~\eqref{eq:translational}--\eqref{eq:rotational}, and the quaternion kinematics~\eqref{eq:quaternion}, the complete coupled system is recast in the first-order state-space form of~\Cref{eq:state_space}. Linearisation about the trim state $\mathbf{w}_0$ yields:
\begin{equation}
\dot{\mathbf{w}} = \mathbf{A}\mathbf{w} + \mathbf{B}_c \mathbf{u}_c + \mathbf{B}_g \mathbf{u}_d + \mathbf{F}_{\text{nl}}(\mathbf{w})
\label{eq:first_order}
\end{equation}
where $\mathbf{A} = \partial\mathbf{R}/\partial\mathbf{w}|_{\mathbf{w}_0} \in \mathbb{R}^{n \times n}$ is the Jacobian matrix of the coupled system at equilibrium, $\mathbf{B}_c$ and $\mathbf{B}_g$ are the input matrices for control and gust, and $\mathbf{F}_{\text{nl}}$ collects all terms beyond the first-order linearisation~\citep{Hesse2014, Tantaroudas2017bookchapter}. The Jacobian has a characteristic block structure reflecting the coupling between subsystems:
\begin{equation}
\mathbf{A} = \begin{bmatrix}
\mathbf{A}_{ff} & \mathbf{A}_{fs} & \mathbf{A}_{fr} \\
\mathbf{A}_{sf} & \mathbf{A}_{ss} & \mathbf{A}_{sr} \\
\mathbf{A}_{rf} & \mathbf{A}_{rs} & \mathbf{A}_{rr}
\end{bmatrix}
\label{eq:jacobian_blocks}
\end{equation}
where the off-diagonal blocks encode the coupling between fluid ($f$), structural ($s$), and rigid-body ($r$) partitions.

\section{Nonlinear Model Order Reduction}
\label{sec:nmor}

The NMOR methodology consists of three steps: (i) Taylor series expansion of the nonlinear residual, (ii) construction of a compact eigenvector basis from the Jacobian eigenspectrum, and (iii) Galerkin projection of the expanded equations onto this basis.

\subsection{Taylor Series Expansion}

The nonlinear residual in~\Cref{eq:state_space} is expanded in a multivariate Taylor series around the equilibrium $\mathbf{w}_0$, retaining terms up to third order~\citep{DaRonch2013control, DaRonch2013gust}:
\begin{equation}
R_k(\mathbf{w}) = \sum_{i=1}^{n} A_{ki} \Delta w_i + \frac{1}{2}\sum_{i,j=1}^{n} B_{kij} \Delta w_i \Delta w_j + \frac{1}{6}\sum_{i,j,l=1}^{n} C_{kijl} \Delta w_i \Delta w_j \Delta w_l + \mathcal{O}(\|\Delta\mathbf{w}\|^4)
\label{eq:taylor_component}
\end{equation}
for each component $k = 1, \ldots, n$, where $\Delta\mathbf{w} = \mathbf{w} - \mathbf{w}_0$. In compact operator notation:
\begin{equation}
\mathbf{R}(\mathbf{w}) \approx \mathbf{A} \Delta\mathbf{w} + \frac{\partial \mathbf{R}}{\partial \mathbf{u}_c} \Delta\mathbf{u}_c + \frac{\partial \mathbf{R}}{\partial \mathbf{u}_d} \Delta\mathbf{u}_d + \frac{1}{2}\mathcal{B}(\Delta\mathbf{w}, \Delta\mathbf{w}) + \frac{1}{6}\mathcal{C}(\Delta\mathbf{w}, \Delta\mathbf{w}, \Delta\mathbf{w})
\label{eq:taylor}
\end{equation}
where $\mathcal{B}: \mathbb{R}^n \times \mathbb{R}^n \to \mathbb{R}^n$ is the symmetric bilinear operator with components $B_{kij} = \partial^2 R_k / \partial w_i \partial w_j |_{\mathbf{w}_0}$, and $\mathcal{C}: \mathbb{R}^n \times \mathbb{R}^n \times \mathbb{R}^n \to \mathbb{R}^n$ is the symmetric trilinear operator with components $C_{kijl} = \partial^3 R_k / \partial w_i \partial w_j \partial w_l |_{\mathbf{w}_0}$.

\subsection{Eigenvector Basis Construction}
\label{sec:basis}

The right generalised eigenvalue problem of the Jacobian is:
\begin{equation}
\mathbf{A} \boldsymbol{\phi}_i = \lambda_i \boldsymbol{\phi}_i, \qquad i = 1, \ldots, n
\end{equation}
and the left eigenvalue problem:
\begin{equation}
\boldsymbol{\psi}_j^H \mathbf{A} = \lambda_j \boldsymbol{\psi}_j^H, \qquad j = 1, \ldots, n
\end{equation}
where $\lambda_i \in \mathbb{C}$ are the eigenvalues and $\boldsymbol{\phi}_i, \boldsymbol{\psi}_i \in \mathbb{C}^n$ are the right and left eigenvectors. The left and right eigenvectors satisfy the biorthonormality condition:
\begin{equation}
\boldsymbol{\Psi}^H \boldsymbol{\Phi} = \mathbf{I}_m
\label{eq:biortho}
\end{equation}
where $\boldsymbol{\Phi} = [\boldsymbol{\phi}_1, \ldots, \boldsymbol{\phi}_m] \in \mathbb{C}^{n \times m}$ and $\boldsymbol{\Psi} = [\boldsymbol{\psi}_1, \ldots, \boldsymbol{\psi}_m] \in \mathbb{C}^{n \times m}$ collect the $m$ selected right and left eigenvectors. This biorthonormality is critical: it ensures that the projection diagonalises the linear part of the dynamics, yielding modal equations that are decoupled at the linear level and coupled only through the nonlinear interaction terms.

A reduced basis of $m \ll n$ eigenvectors is selected according to the following physical strategy:
\begin{enumerate}
\item \textbf{Real eigenvalues near the origin}: These correspond to rigid-body dynamics (phugoid, spiral, Dutch roll at low damping) and to gust penetration coupling modes. They are essential for capturing the mean response and slow dynamics of the aircraft centre of mass.
\item \textbf{Lightly damped complex eigenvalues}: These correspond to the dominant structural modes (first bending, second bending, first torsion, etc.)\ modified by aerodynamic damping. They govern the oscillatory aeroelastic response and are selected in order of increasing damping ratio $\zeta_i = -\text{Re}(\lambda_i) / |\lambda_i|$.
\end{enumerate}
An example of this basis selection strategy is presented in~\Cref{fig:basis_3dof} for the three-DOF aerofoil test case.

The state perturbation is then approximated by the truncated eigenexpansion:
\begin{equation}
\Delta\mathbf{w} \approx \boldsymbol{\Phi} \mathbf{z} + \bar{\boldsymbol{\Phi}} \bar{\mathbf{z}}
\label{eq:projection}
\end{equation}
where $\mathbf{z} \in \mathbb{C}^m$ is the reduced state vector and the overbar denotes complex conjugation.

\subsection{Reduced-Order Equations}

Substituting the expansion~\eqref{eq:projection} into the Taylor series~\eqref{eq:taylor} and premultiplying by $\boldsymbol{\Psi}^H$ (Galerkin projection using the left eigenvectors), the reduced-order model is obtained:
\begin{equation}
\frac{dz_k}{dt} = \lambda_k z_k + \sum_{i,j=1}^{m} D_{kij} z_i z_j + \sum_{i,j,l=1}^{m} E_{kijl} z_i z_j z_l + \boldsymbol{\psi}_k^H \mathbf{B}_c \Delta\mathbf{u}_c + \boldsymbol{\psi}_k^H \mathbf{B}_g \Delta\mathbf{u}_d
\label{eq:rom}
\end{equation}
for $k = 1, \ldots, m$, where the \emph{second-order interaction coefficients} are:
\begin{equation}
D_{kij} = \frac{1}{2} \boldsymbol{\psi}_k^H \mathcal{B}(\boldsymbol{\phi}_i, \boldsymbol{\phi}_j)
\label{eq:D_coeff}
\end{equation}
and the \emph{third-order interaction coefficients} are:
\begin{equation}
E_{kijl} = \frac{1}{6} \boldsymbol{\psi}_k^H \mathcal{C}(\boldsymbol{\phi}_i, \boldsymbol{\phi}_j, \boldsymbol{\phi}_l)
\label{eq:E_coeff}
\end{equation}

The reduced system~\eqref{eq:rom} has dimension $m$ (typically 4--9), compared to $n$ (typically 500--2,000) for the full-order model. The reduced model retains physical nonlinear coupling between modes and is \emph{excitation-independent}: once the coefficients $D_{kij}$ and $E_{kijl}$ are computed, the ROM can be excited by arbitrary gust profiles, control inputs, or initial conditions without regeneration~\citep{Tantaroudas2017bookchapter, Tantaroudas2015scitech}. Furthermore, the framework is formulation-independent, requiring only the ability to evaluate the residual $\mathbf{R}(\mathbf{w})$ and solve the eigenvalue problem for $\mathbf{A}$, which makes it readily applicable to higher-fidelity aerodynamic models.

\subsection{Matrix-Free Computation of Higher-Order Terms}
\label{sec:matrixfree}

Explicit formation of the bilinear operator $\mathcal{B} \in \mathbb{R}^{n \times n \times n}$ or the trilinear operator $\mathcal{C} \in \mathbb{R}^{n \times n \times n \times n}$ is prohibitive for large systems. Instead, the action of these operators on specific eigenvector directions is computed via matrix-free finite differences~\citep{DaRonch2013control}.

The bilinear operator acting on two eigenvectors $\boldsymbol{\phi}_i$ and $\boldsymbol{\phi}_j$ is approximated by:
\begin{equation}
\mathcal{B}(\boldsymbol{\phi}_i, \boldsymbol{\phi}_j) \approx \frac{\mathbf{R}(\mathbf{w}_0 + \epsilon\boldsymbol{\phi}_i + \epsilon\boldsymbol{\phi}_j) - \mathbf{R}(\mathbf{w}_0 + \epsilon\boldsymbol{\phi}_i) - \mathbf{R}(\mathbf{w}_0 + \epsilon\boldsymbol{\phi}_j) + \mathbf{R}(\mathbf{w}_0)}{\epsilon^2}
\label{eq:B_fd}
\end{equation}
where $\epsilon$ is a perturbation parameter chosen to balance truncation and round-off errors~\citep{DaRonch2013control}.

Similarly, the trilinear operator is computed via third-order finite differences:
\begin{multline}
\mathcal{C}(\boldsymbol{\phi}_i, \boldsymbol{\phi}_j, \boldsymbol{\phi}_l) \approx \frac{1}{\epsilon^3} \Big[ \mathbf{R}(\mathbf{w}_0 + \epsilon\boldsymbol{\phi}_i + \epsilon\boldsymbol{\phi}_j + \epsilon\boldsymbol{\phi}_l) \\
- \mathbf{R}(\mathbf{w}_0 + \epsilon\boldsymbol{\phi}_i + \epsilon\boldsymbol{\phi}_j) - \mathbf{R}(\mathbf{w}_0 + \epsilon\boldsymbol{\phi}_i + \epsilon\boldsymbol{\phi}_l) - \mathbf{R}(\mathbf{w}_0 + \epsilon\boldsymbol{\phi}_j + \epsilon\boldsymbol{\phi}_l) \\
+ \mathbf{R}(\mathbf{w}_0 + \epsilon\boldsymbol{\phi}_i) + \mathbf{R}(\mathbf{w}_0 + \epsilon\boldsymbol{\phi}_j) + \mathbf{R}(\mathbf{w}_0 + \epsilon\boldsymbol{\phi}_l) - \mathbf{R}(\mathbf{w}_0) \Big]
\label{eq:C_fd}
\end{multline}

The computational cost of building the ROM is dominated by the residual evaluations: the second-order coefficients require $\mathcal{O}(m^2)$ evaluations, and the third-order coefficients require $\mathcal{O}(m^3)$ evaluations. For $m = 9$ modes, this amounts to approximately 81 (second-order) and 729 (third-order) residual evaluations. Since each residual evaluation involves a single pass through the full-order equations, the ROM construction cost is a modest multiple of the cost of the trim solution, and is incurred only once. The ROM can subsequently be time-marched at negligible cost compared to the FOM.

\section{Results}
\label{sec:results}

The NMOR framework is validated on three test cases of progressively increasing complexity, summarised in~\Cref{tab:test_cases}. These test cases were previously employed in the published studies~\citep{Tantaroudas2015scitech, Tantaroudas2017bookchapter, DaRonch2013gust} and are reproduced here to provide a self-contained presentation.

\begin{table}[htbp]
\centering
\caption{Summary of test cases for NMOR validation.}
\label{tab:test_cases}
\begin{tabular}{@{}lccc@{}}
\toprule
Property & 3-DOF Aerofoil & HALE Aircraft & VFA Flying-Wing \\
\midrule
Full-order DOFs ($n$) & 14 & 2,016 & 1,616 \\
Structural states & 6 & 1,200 & 960 \\
Aerodynamic states & 8 & 800 & 640 \\
Rigid-body + quaternion & --- & 16 & 16 \\
Retained modes ($m$) & 4 & 9 & 9 \\
Reduction ratio & 3.5:1 & 224:1 & 180:1 \\
Typical speedup & modest & significant & up to 600$\times$ \\
\bottomrule
\end{tabular}
\end{table}

\subsection{Three-Degree-of-Freedom Aerofoil}
\label{sec:3dof}

\begin{figure}[htbp]
\centering
\includegraphics[width=0.65\textwidth]{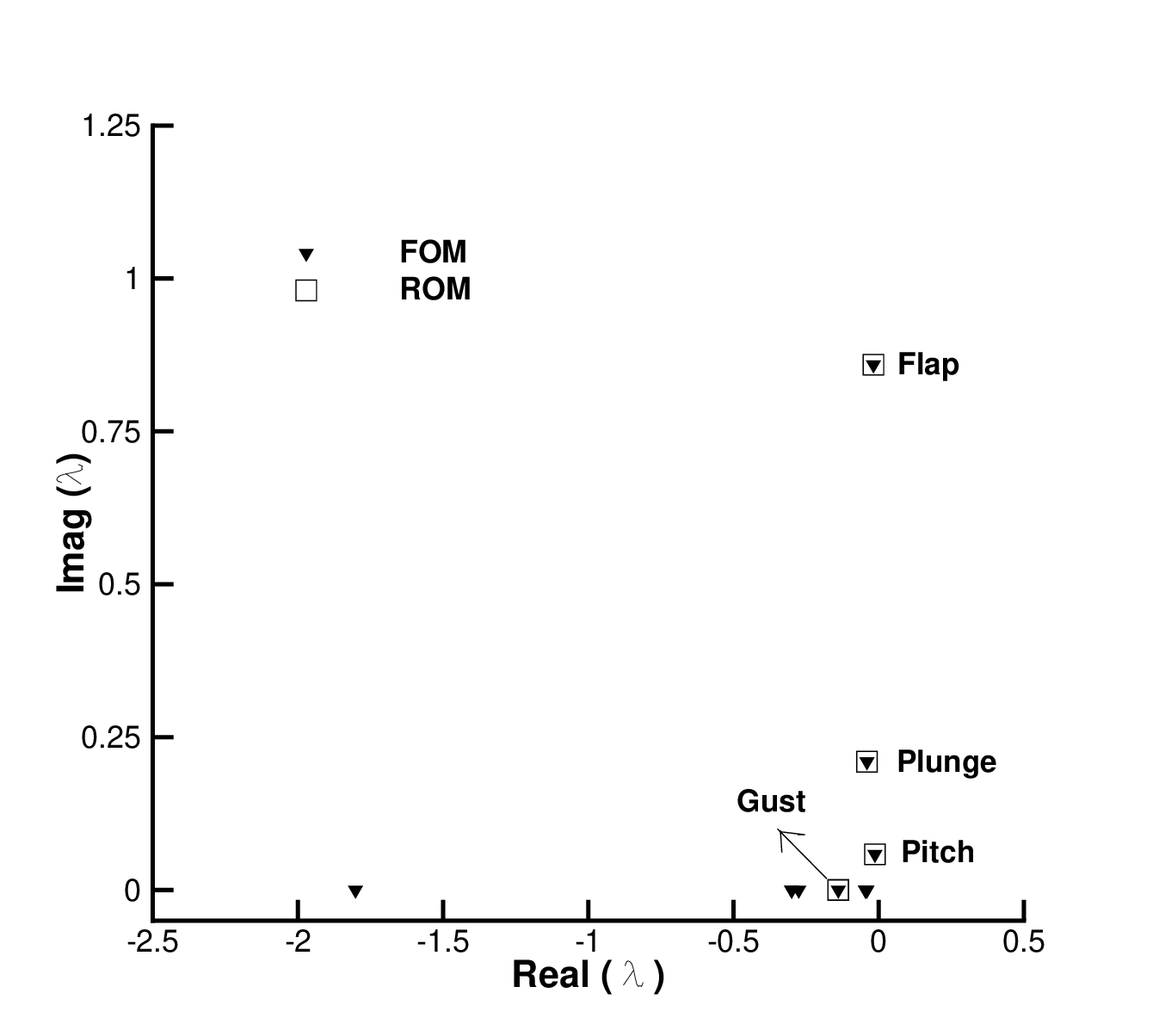}
\caption{Eigenspectrum example of a three-DOF aerofoil system at $U^* = 0.9 U_L^*$ showing the selected eigenvalues (circled) that form the reduced basis. Real eigenvalues near the origin capture gust coupling, while lightly damped complex pairs correspond to pitch, plunge, and flap modes.}
\label{fig:basis_3dof}
\end{figure}

\begin{figure}[htbp]
\centering
\includegraphics[width=0.7\textwidth]{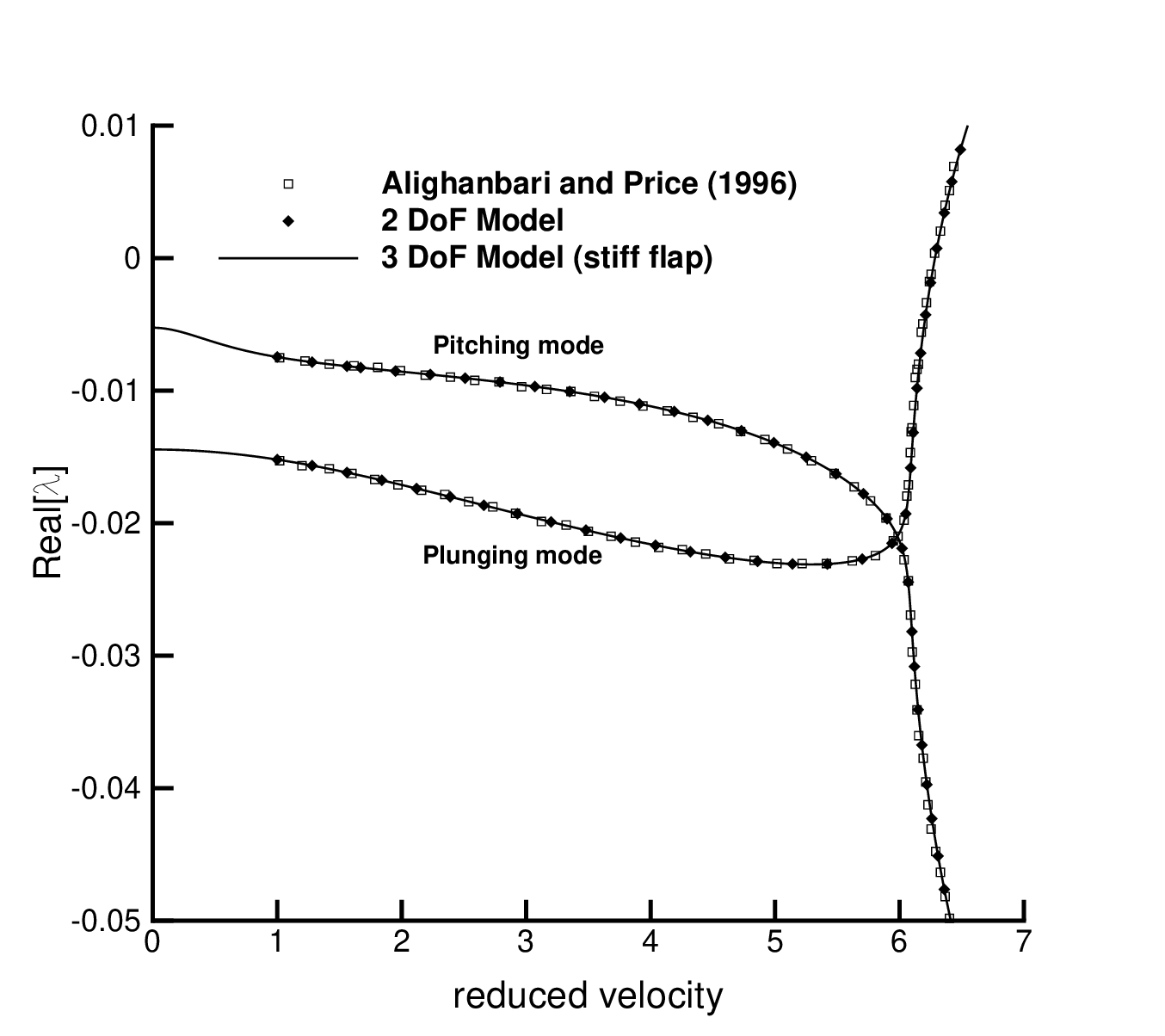}
\caption{Eigenvalue mode trace (real part) of the three-DOF aerofoil as a function of non-dimensional freestream velocity $U^*$ for Case~1~\citep{Alighanbari1995}. The merging of the pitch and plunge branches indicates the onset of coupled-mode (binary) flutter at $U^*_L = 6.285$. Results from the present 3-DOF model (solid line) and a 2-DOF model (filled diamonds) are compared with the reference solution of Alighanbari and Price (open squares).}
\label{fig:modetrace_3dof}
\end{figure}

The first test case is a two-dimensional aerofoil section with three degrees of freedom, pitch ($\alpha$), plunge ($\xi$), and trailing-edge flap deflection ($\delta$), with unsteady aerodynamics based on Theodorsen's theory~\citep{Theodorsen1935}. The structural model incorporates cubic hardening nonlinearities in the pitch and plunge restoring forces:
\begin{align}
M_\alpha &= K_\alpha \alpha + K_{\alpha_3} \alpha^3 \label{eq:cubic_pitch}\\
F_\xi &= K_\xi \xi + K_{\xi_3} \xi^3 \label{eq:cubic_plunge}
\end{align}
with $K_{\alpha_3} = 3$ and $K_{\xi_3} = 1.0$ (non-dimensionalised). The eigenspectrum of this system and the selected eigenvector basis are shown in~\Cref{fig:basis_3dof}, while the eigenvalue mode trace as a function of freestream velocity is presented in~\Cref{fig:modetrace_3dof}.

The full-order model consists of 14 states: 8 augmented aerodynamic states from the Wagner and K\"ussner function approximations (two per indicial function per degree of freedom), and 6 structural states (generalised displacements $\alpha, \xi, \delta$ and their time derivatives).

Two parameter sets are considered for validation. Case~1~\citep{Alighanbari1995} uses a linear flutter speed $U_L^* = 6.285$, mass ratio $\mu = 100$, frequency ratio $\bar{\omega} = 0.2$, and elastic axis at $a = -0.5$. Case~2~\citep{Irani2011} uses $U_L^* = 4.663$, $\mu = 100$, and $\bar{\omega}_1 = 1.2$.

The ROM retains 4 eigenvalues: 3 complex pairs corresponding to the structural modes (pitch, plunge, flap) and 1 real eigenvalue associated with the K\"ussner gust coupling. Validation is performed in both sub-critical and post-flutter regimes.

\subsubsection{Sub-critical response}

\begin{figure}[htbp]
\centering
\begin{subfigure}[b]{0.48\textwidth}
\includegraphics[width=\textwidth]{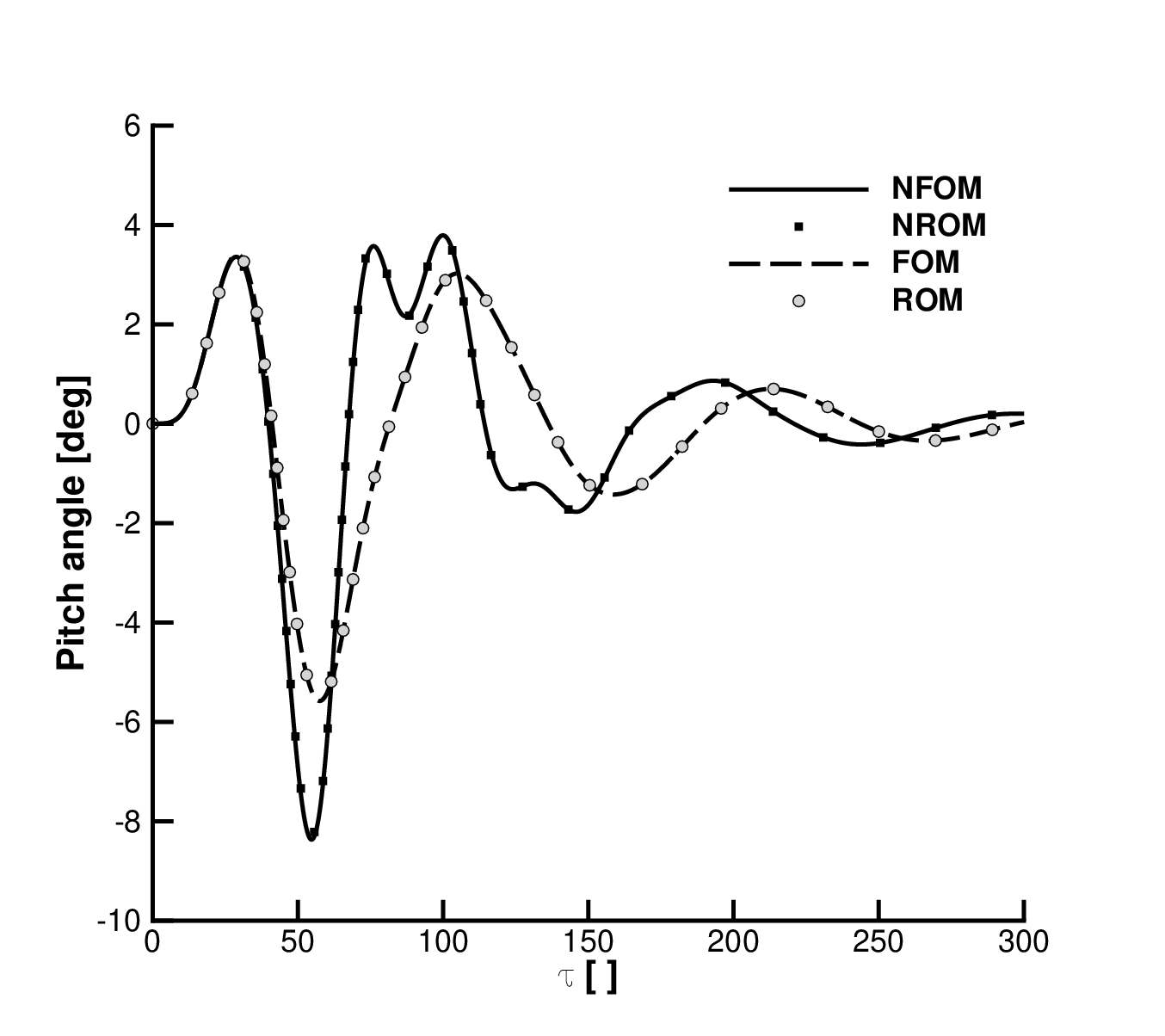}
\caption{Pitch response $\alpha(t)$}
\label{fig:nrom_3dof_pitch}
\end{subfigure}
\hfill
\begin{subfigure}[b]{0.48\textwidth}
\includegraphics[width=\textwidth]{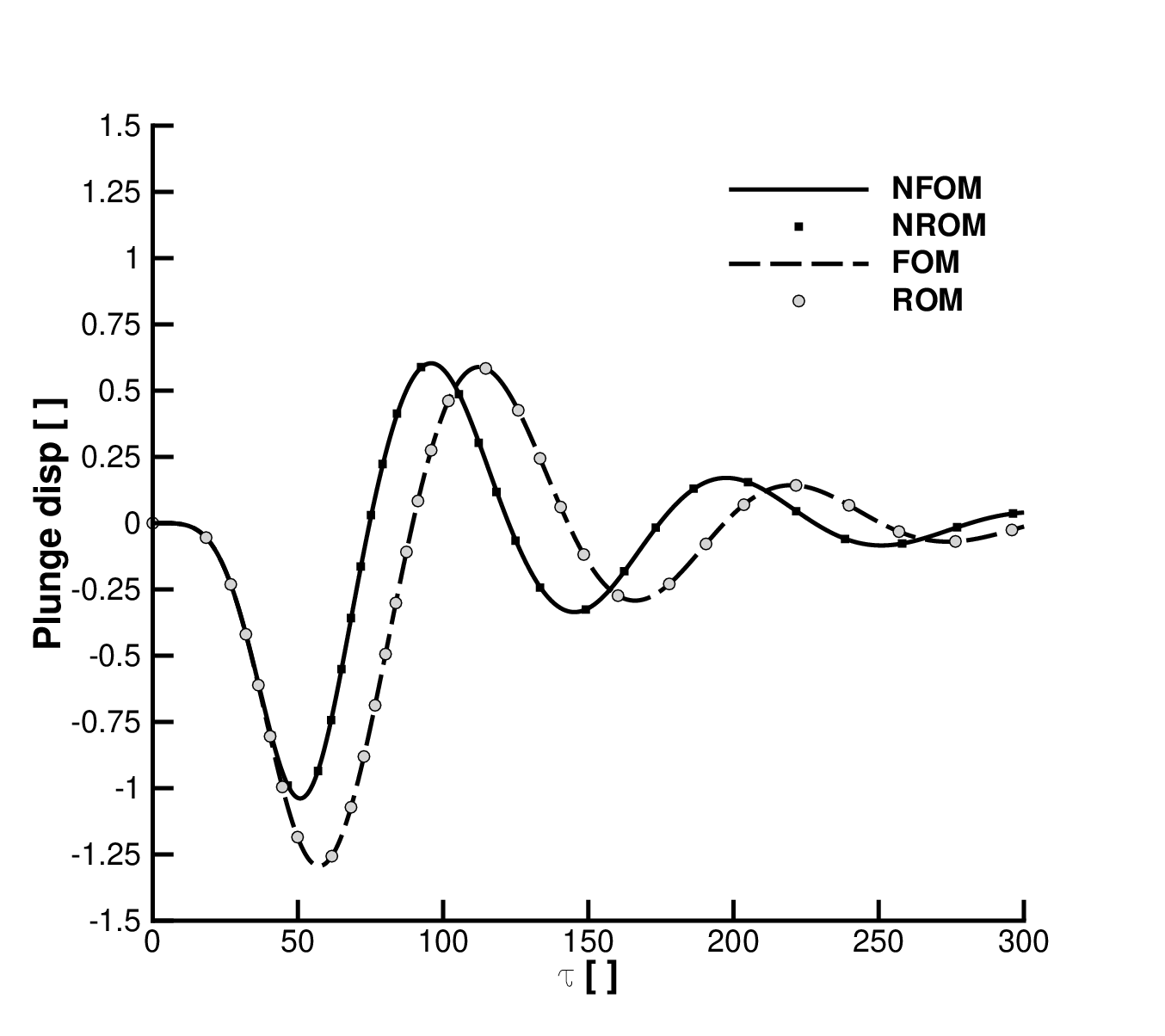}
\caption{Plunge response $\xi(t)$}
\label{fig:nrom_3dof_plunge}
\end{subfigure}
\caption{Time histories of the three-DOF aerofoil at sub-critical conditions ($U^* = 0.9 U_L^*$, Case~1): nonlinear full-order model (NFOM, solid), nonlinear ROM (NROM, filled squares), linear full-order model (FOM, dashed), and linear ROM (ROM, open circles). The nonlinear ROM with 4 retained modes reproduces the full-model decay to within plotting accuracy.}
\label{fig:nrom_subcritical}
\end{figure}

\begin{figure}[htbp]
\centering
\includegraphics[width=0.55\textwidth]{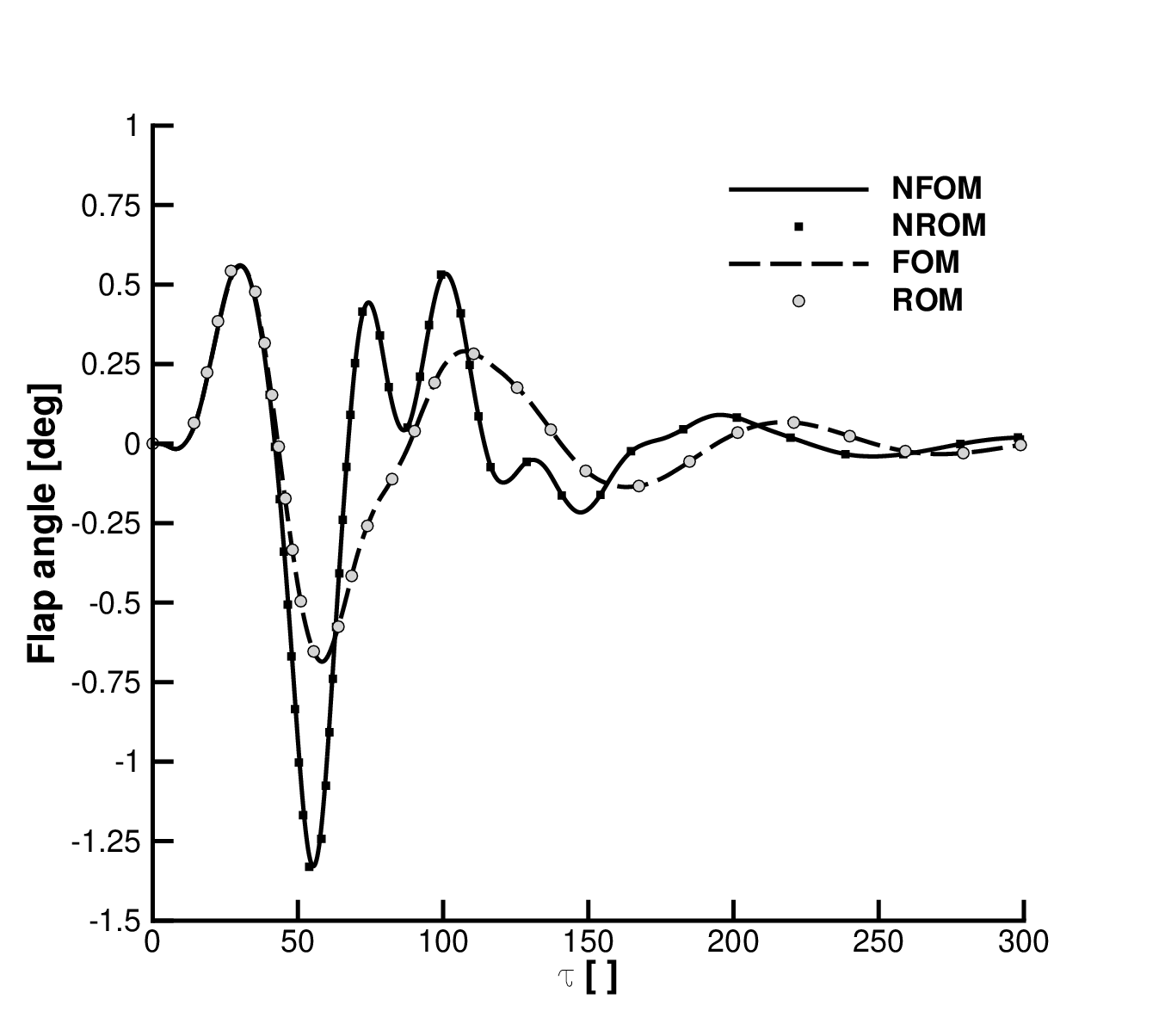}
\caption{Flap deflection response $\delta(t)$ at sub-critical conditions ($U^* = 0.9 U_L^*$, Case~1): nonlinear full-order model (NFOM, solid), nonlinear ROM (NROM, filled squares), linear full-order model (FOM, dashed), and linear ROM (ROM, open circles).}
\label{fig:nrom_3dof_delta}
\end{figure}

At $U^* = 0.9 U_L^*$, the system is stable and decays to the trivial equilibrium. The NMOR reproduces the full-order time histories for pitch, plunge, and flap with negligible error, as shown in~\Cref{fig:nrom_3dof_pitch,fig:nrom_3dof_plunge,fig:nrom_3dof_delta}. In this regime, the cubic hardening nonlinearity plays a minor role, and the linear ROM also performs well. The nonlinear ROM achieves a modest computational speedup per simulation owing to the small size of the full-order model.

\subsubsection{Post-flutter limit-cycle oscillation}

Above the linear flutter speed ($U^* > U_L^*$), the cubic hardening nonlinearity limits the growing oscillation to a stable limit-cycle oscillation (LCO). The LCO amplitude depends on the balance between the linear instability growth rate and the nonlinear restoring force.

The nonlinear ROM captures the post-flutter limit-cycle behaviour, confirming that the second-order Taylor expansion of the residual, which produces quadratic interaction terms in the reduced equations, is sufficient to reproduce the cubic nonlinear behaviour of the structural restoring forces. The LCO amplitude predicted by the nonlinear ROM closely matches the full-order model, demonstrating the fidelity of the reduced model in capturing nonlinear dynamics~\citep{DaRonch2013control, Tantaroudas2017bookchapter}. The linear ROM, by contrast, lacks the stabilising cubic terms and cannot predict bounded post-flutter oscillations~\citep{DaRonch2013control}.

\subsection{HALE Aircraft Configuration}
\label{sec:hale}

\begin{figure}[htbp]
\centering
\includegraphics[width=0.7\textwidth]{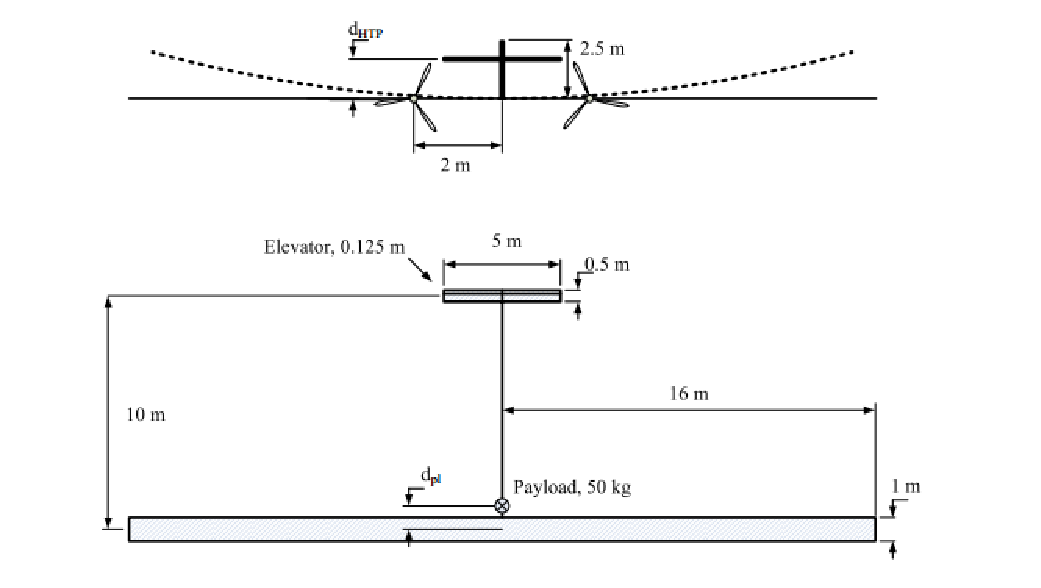}
\caption{Three-view schematic of the HALE aircraft configuration used for NMOR validation. The aircraft has a 32-m wingspan (16-m semi-span), a rigid fuselage, horizontal and vertical tail surfaces, and two propulsive units mounted on the wing. Properties follow~\citet{Patil2001} with modifications from~\citet{Hesse2014}.}
\label{fig:hale_geometry}
\end{figure}

The second test case is a representative HALE aircraft configuration with geometric properties closely matching the benchmark model of~\citet{Patil2001}. The aircraft layout and geometry are illustrated in~\Cref{fig:hale_geometry}.

The main wing has a 32-m wingspan with 1-m chord, mass per unit length of 0.75~kg/m, bending stiffness $EI_2 = 2\sigma_1 \times 10^4$~Nm$^2$, torsional stiffness $GJ = \sigma_1 \times 10^4$~Nm$^2$, and chordwise bending stiffness $EI_3 = \sigma_2 \times 10^6$~Nm$^2$, where $\sigma_1$ and $\sigma_2$ are non-dimensional flexibility parameters. Unless otherwise stated, the nominal values $\sigma_1 = 1$ and $\sigma_2 = 5$ are used, giving $EI_2 = 2 \times 10^4$~Nm$^2$ and $GJ = 1 \times 10^4$~Nm$^2$. The aircraft carries a 50-kg payload on the fuselage and is powered by two propellers mounted on the wing, providing sufficient thrust for level flight at $U = 25$~m/s at an altitude of 20,000~m (air density $\rho = 0.0889$~kg/m$^3$).

A convergence study confirmed that a finite element mesh of 100 nodes provides excellent agreement with published data~\citep{Patil2001}. The resulting full-order model has $n = 2{,}016$ states: 1,200 structural states (6 displacement and 6 velocity degrees of freedom per node in first-order form), 12 rigid-body states ($\mathbf{v}_B$, $\boldsymbol{\omega}_B$), 4 quaternion components, and 800 augmented aerodynamic states (8 per strip).

\subsubsection{Structural validation}

The linearised structural natural frequencies are validated against published results in~\Cref{tab:hale_freq}.

\begin{table}[htbp]
\centering
\caption{Comparison of HALE aircraft natural frequencies (rad/s) at zero freestream velocity.}
\label{tab:hale_freq}
\begin{tabular}{@{}lcc@{}}
\toprule
Mode & Present & Patil et al.\ \citeyearpar{Patil2001} \\
\midrule
1st bending & 2.24 & 2.24 \\
2nd bending & 14.07 & 14.60 \\
1st torsion & 31.04 & 31.14 \\
1st in-plane & 31.71 & 31.73 \\
3rd bending & 39.52 & 44.01 \\
\bottomrule
\end{tabular}
\end{table}

The excellent agreement confirms the structural model implementation. The flutter speed comparison is reported in~\Cref{tab:hale_flutter}.

\begin{table}[htbp]
\centering
\caption{Flutter speed comparison for the cantilever beam.}
\label{tab:hale_flutter}
\begin{tabular}{@{}lcc@{}}
\toprule
 & $U_L$ [m/s] & $\omega$ [rad/s] \\
\midrule
Present analysis & 31.2 & 22.1 \\
Murua et al.~\citep{Murua2012} & 33.0 & 22.0 \\
Patil et al.~\citep{Patil2001} & 32.2 & 22.6 \\
\bottomrule
\end{tabular}
\end{table}

The strip theory assumption produces flutter speeds closer to the results of~\citet{Patil2001}, who employed two-dimensional Peters aerodynamic modelling. Minor differences with the UVLM results are attributed to three-dimensional aerodynamic effects.

\subsubsection{Static aeroelastic deformation}

\begin{figure}[htbp]
\centering
\includegraphics[width=0.6\textwidth]{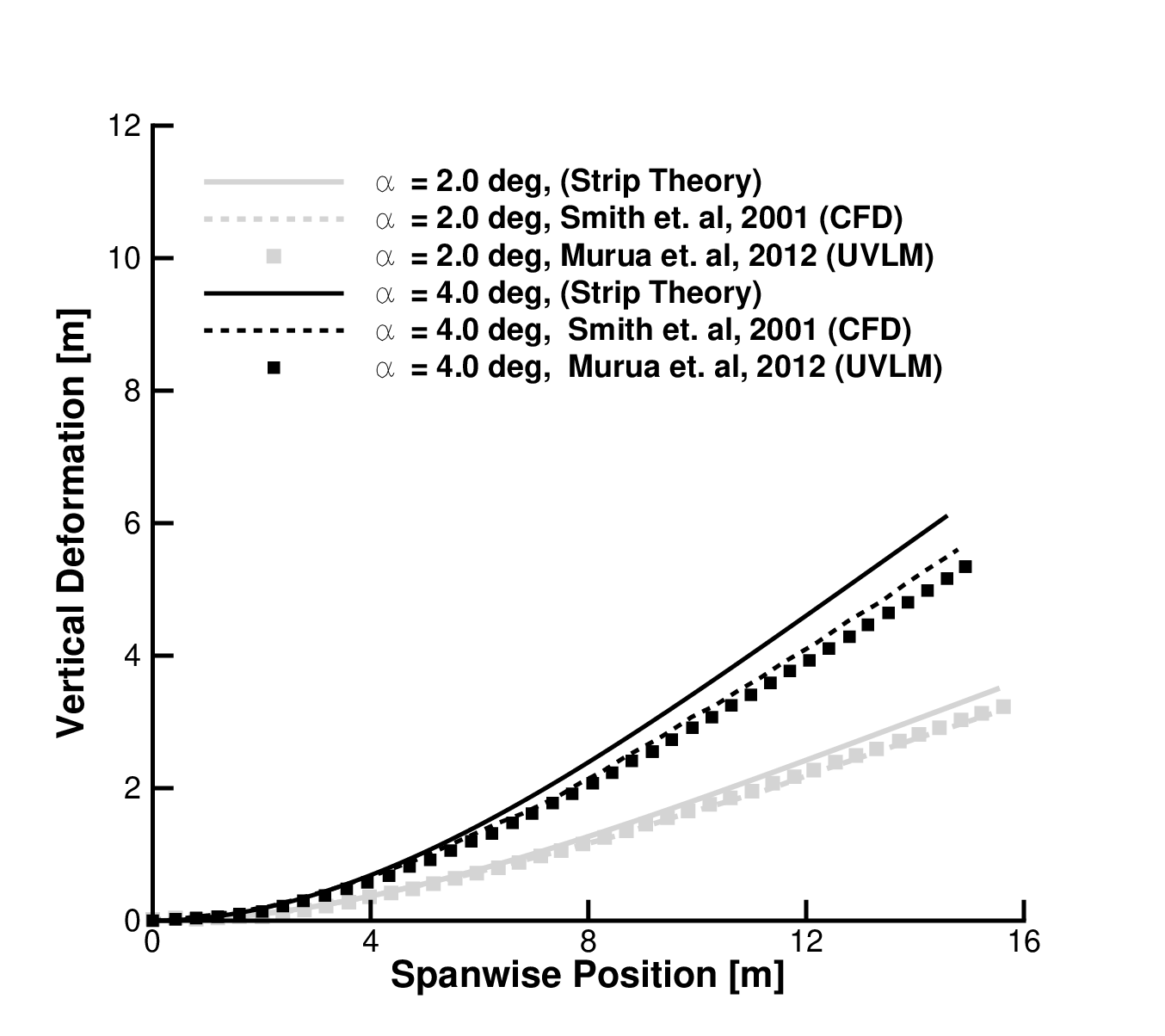}
\caption{Static aeroelastic wing-tip vertical deflection of the HALE aircraft as a function of spanwise position at two angles of attack ($\alpha = 2^\circ$ and $\alpha = 4^\circ$): present strip-theory model (lines), UVLM from Murua et al.~\citep{Murua2012} (squares), and 3D Euler CFD from Smith et al.\ (dashed). Agreement is good at $\alpha = 2^\circ$; divergence at higher incidence reflects the limitations of two-dimensional aerodynamics.}
\label{fig:hale_static}
\end{figure}

\Cref{fig:hale_static} compares the static aeroelastic wing deformation at the trim condition with results from a three-dimensional Euler CFD solver and UVLM. At moderate angles of attack ($\alpha = 2^\circ$), the present strip-theory-based model agrees well with these higher-fidelity solutions. At $\alpha = 4^\circ$, three-dimensional aerodynamic effects not captured by strip theory cause the predictions to diverge, as discussed in~\citep{DaRonch2014scitech_flight}.

The trim condition at the design flight speed results in significant wing-tip deflection, confirming that geometric nonlinearities are relevant and must be accounted for in the ROM.

\subsubsection{Vertical equilibrium trimming}
\label{sec:trim}

\begin{figure}[htbp]
\centering
\includegraphics[width=0.45\textwidth]{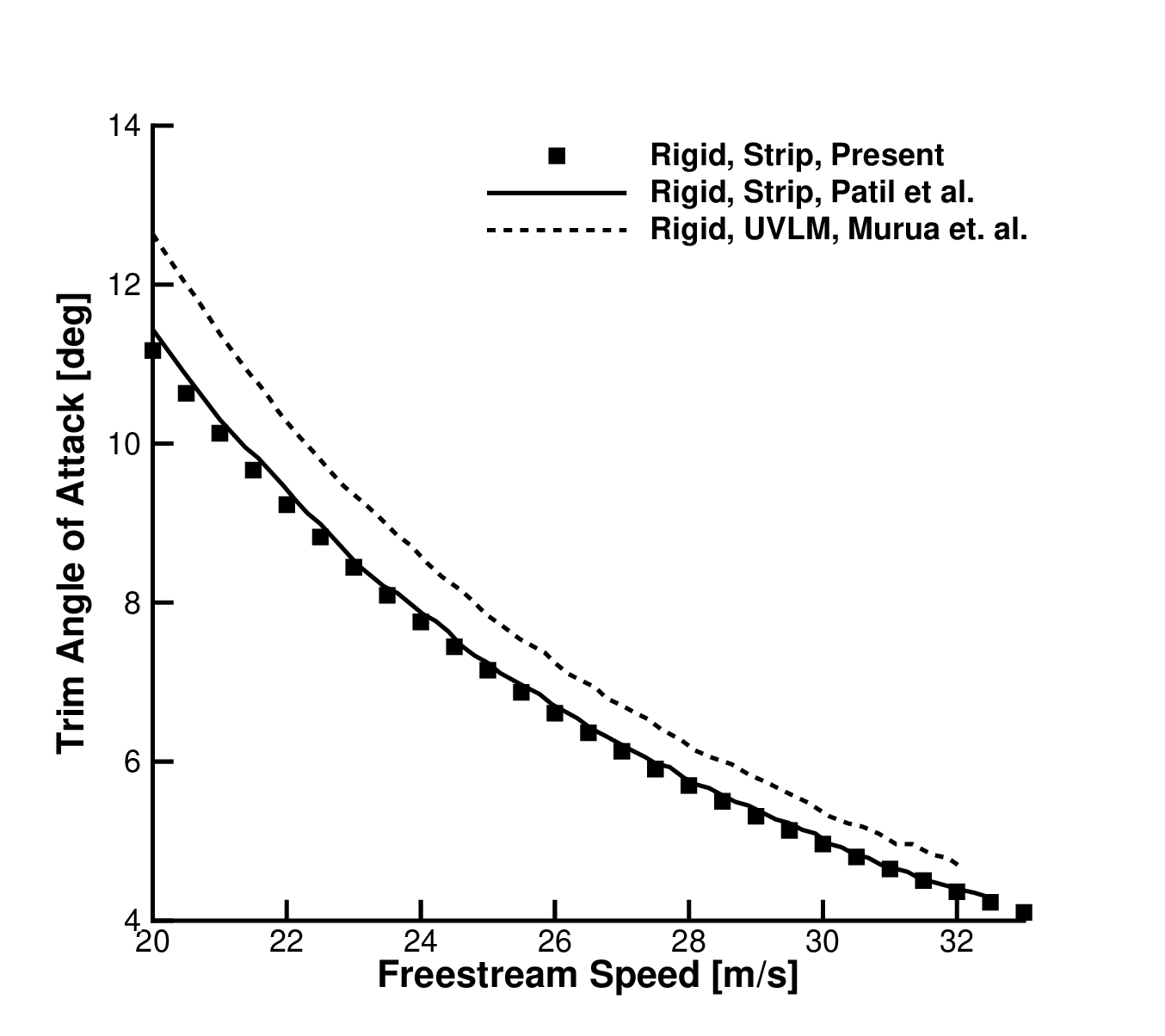}
\includegraphics[width=0.45\textwidth]{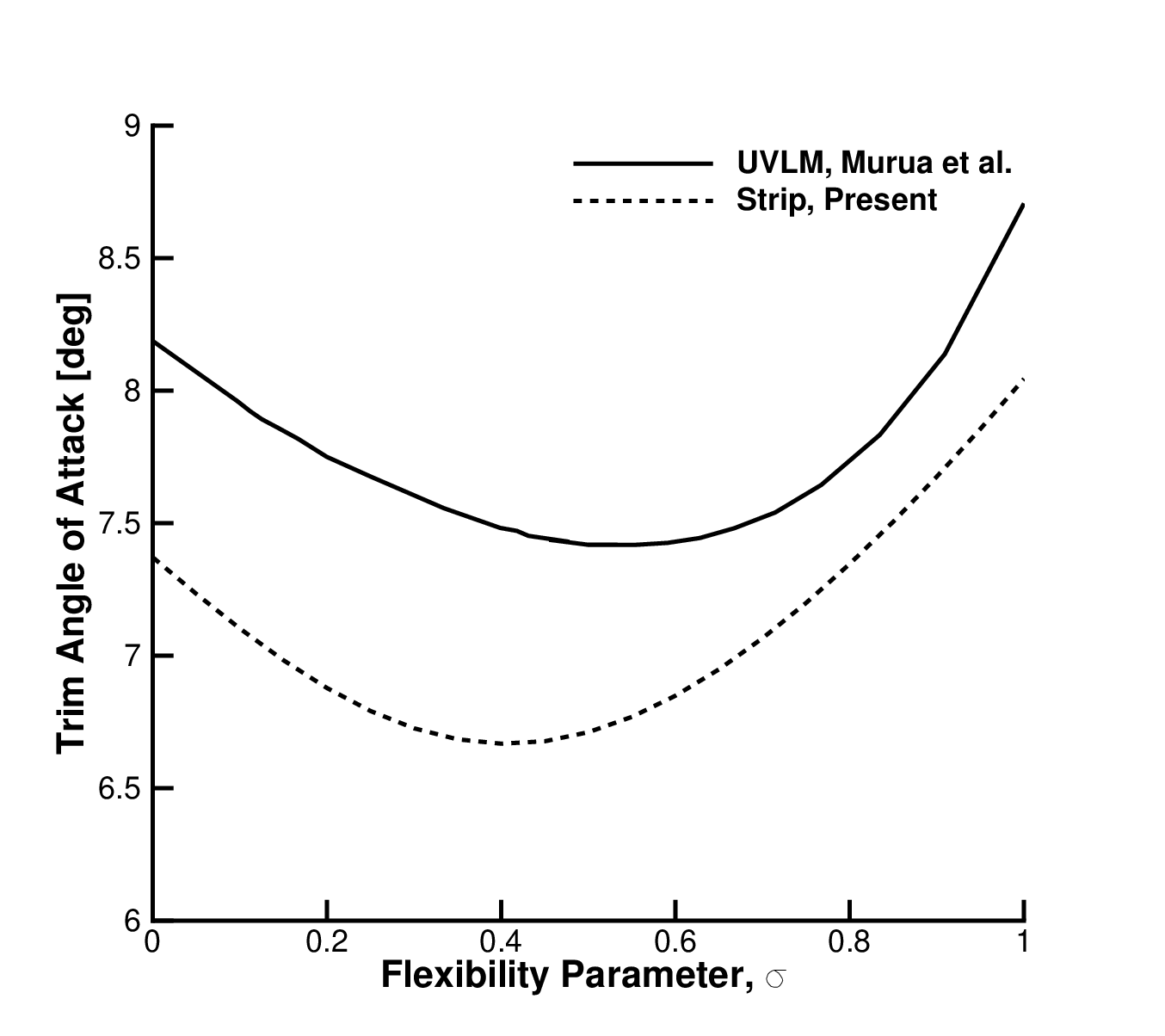}
\caption{Variation of angle of attack with flight speed for vertical force equilibrium ($\sigma_1=1$, $\sigma_2=2$, $d_{pl}=2$, $d_{HTP}=0$). Current results compared to Murua et al.~\citep{Murua2012} and Patil et al.~\citep{Patil2001}.}
\label{fig:hale_trim}
\end{figure}

The rigid-body degrees of freedom introduce additional aerodynamic forces and the first step towards verification of a flying wing is to investigate these effects by trimming the aircraft. The angle of attack needed to counterbalance gravitational forces at different freestream speeds was computed. The payload was placed at a distance $d_{pl}=2$~m. \Cref{fig:hale_trim} shows that the flexible wing gives different predictions from the rigid wing even without control surface rotation. Furthermore, the current strip theory gives the same predictions as with other two-dimensional potential aerodynamics, while UVLM provides a more accurate prediction.

When deformations and bending are very large, a larger angle of attack is needed for vertical force equilibrium because the lift force does not act in the vertical direction when the wing exhibits large nonlinear deformations. The deformed configuration of the trimmed aircraft at 25~m/s is shown in~\Cref{fig:hale_deformation}, in comparison with published results from~\citet{Patil2001}. The two-dimensional aerodynamics provide the same description of the physics of flexible aircraft undergoing large deformations and rigid-body motion compared to previously published data.

\subsubsection{Reduced-order model}

\begin{figure}[htbp]
\centering
\includegraphics[width=0.6\textwidth]{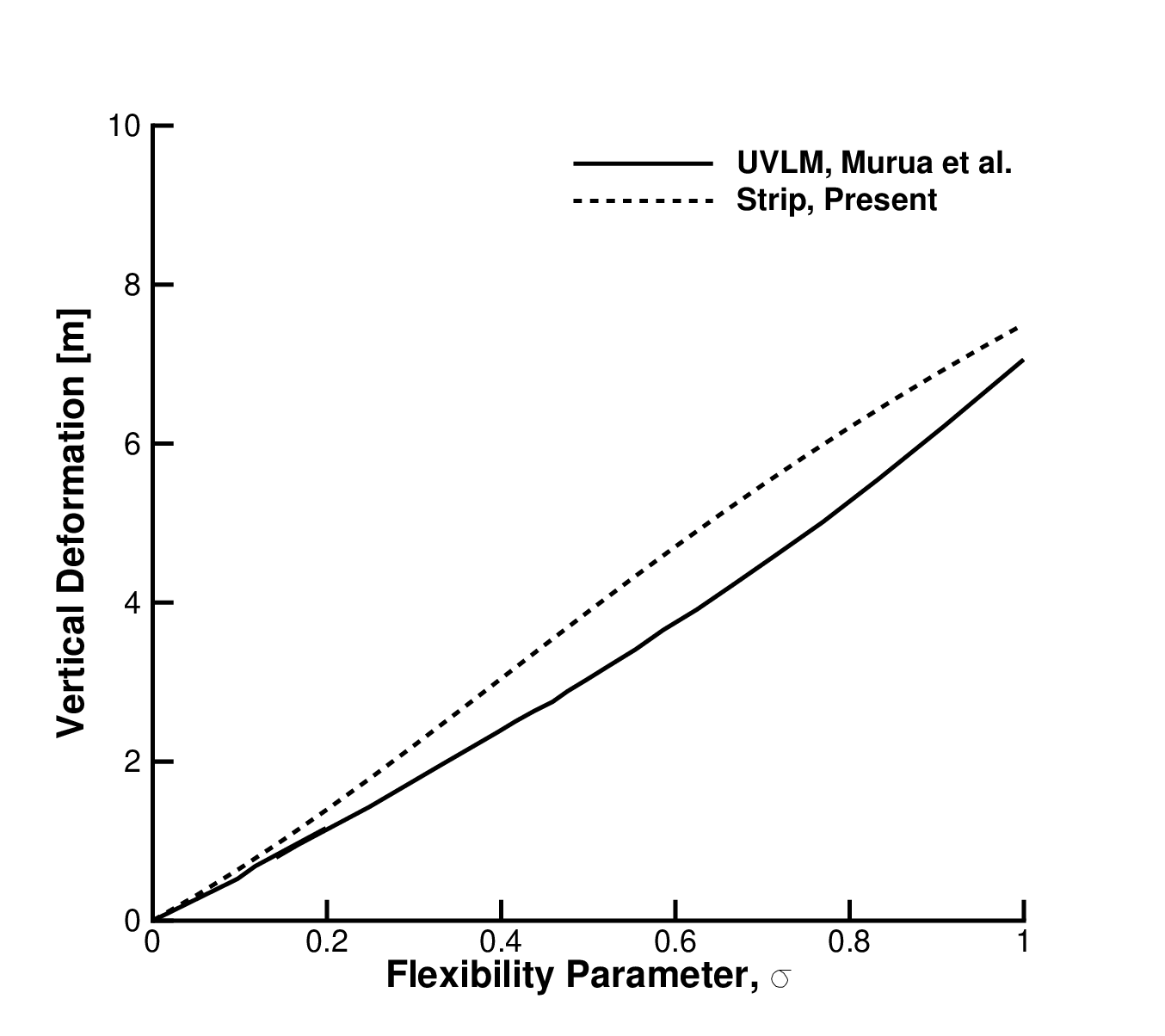}
\caption{HALE aircraft wing-tip vertical deformation as a function of the flexibility parameter $\sigma$: present strip-theory model (dashed) and UVLM from~\citet{Murua2012} (solid). The nonlinear growth of deformation with increasing flexibility confirms the relevance of geometric nonlinearities for very flexible configurations.}
\label{fig:hale_deformation}
\end{figure}

The ROM retains 9 eigenvalues: 4 real eigenvalues near the origin (associated with rigid-body and gust coupling dynamics) and 5 complex conjugate pairs (first through fourth bending and first torsion). The sensitivity of the wing-tip vertical deformation to the flexibility parameter $\sigma$ is shown in~\Cref{fig:hale_deformation}, comparing the present strip-theory model with the UVLM results of~\citet{Murua2012}. As $\sigma$ increases (softer wing), the tip deflection grows nonlinearly, confirming the importance of geometric nonlinearities. The equilibrium deformation at the design flight condition serves as the reference state for the Taylor series expansion in the NMOR procedure.

\subsection{Very Flexible Flying-Wing}
\label{sec:vfa}

The third test case is a 32-m span very flexible free-flying wing, a symmetric, all-wing configuration without fuselage or tail. The aircraft is modelled as a free-flying vehicle with full rigid-body degrees of freedom (12 rigid-body states and 4 quaternion components for attitude propagation), coupled with the flexible structural and unsteady aerodynamic subsystems. This configuration represents the most challenging scenario for NMOR due to the large structural deformations, the coupling between flexibility and rigid-body flight dynamics, and the absence of inherent pitch stiffness that a conventional tail would provide. The structural and aerodynamic properties are summarised in~\Cref{tab:vfa_props}.

\begin{table}[htbp]
\centering
\caption{Properties of the very flexible flying-wing test case.}
\label{tab:vfa_props}
\begin{tabular}{@{}ll@{}}
\toprule
Property & Value \\
\midrule
Wing span & 32 m \\
Chord & 1 m \\
Mass per unit length & 10 kg/m \\
Moment of inertia & 10 kg\,m \\
Torsional stiffness $GJ$ & $1.25 \times 10^4$ Nm$^2$ \\
Bending stiffness $EI_2$ & $2.5 \times 10^4$ Nm$^2$ \\
In-plane stiffness $EI_3$ & $6.25 \times 10^6$ Nm$^2$ \\
Elastic axis & 25\% chord \\
Control surface & 10\% chord trailing-edge flap \\
Altitude & 20,000 m \\
Freestream velocity & 25 m/s \\
Air density & 0.0889 kg/m$^3$ \\
\bottomrule
\end{tabular}
\end{table}

The FE model uses 80 two-noded beam elements (40 per semi-span), yielding $n = 1{,}616$ DOFs: 960 structural, 640 aerodynamic, 12 rigid-body, and 4 quaternion states. The normal modes about the undeformed configuration for the clamped structural model are reported in~\Cref{tab:vfa_freq}.

\begin{table}[htbp]
\centering
\caption{Natural frequencies of the very flexible flying-wing (rad/s).}
\label{tab:vfa_freq}
\begin{tabular}{@{}lc@{}}
\toprule
Mode & Frequency \\
\midrule
1st bending & 4.28 \\
1st torsion & 10.41 \\
1st in-plane & 10.84 \\
2nd bending & 11.92 \\
3rd bending & 17.38 \\
\bottomrule
\end{tabular}
\end{table}

\subsubsection{Eigenvector basis selection}

\begin{figure}[htbp]
\centering
\includegraphics[width=0.6\textwidth]{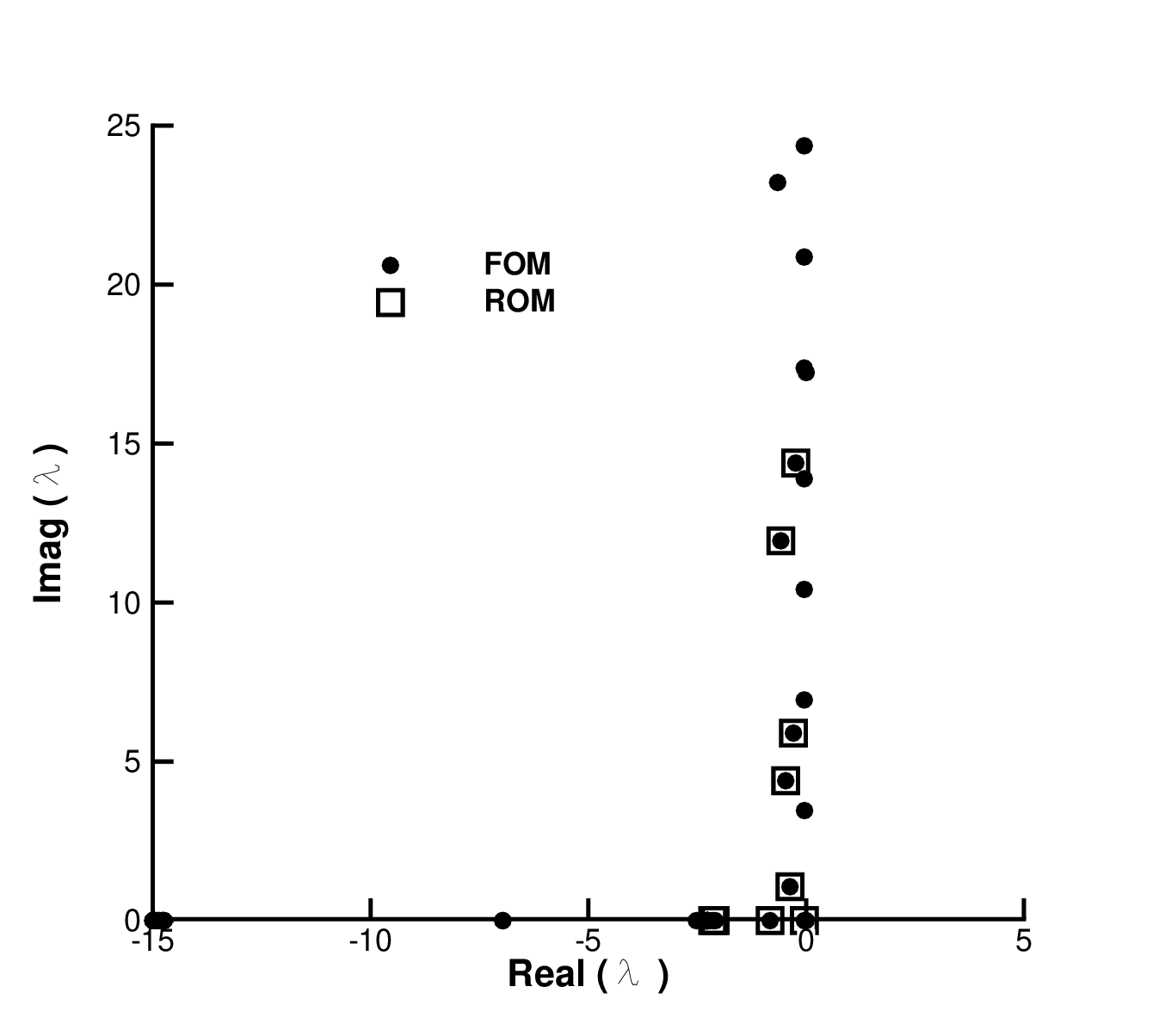}
\caption{Eigenspectrum of the coupled Jacobian for the very flexible flying-wing at the trimmed flight condition. Selected eigenvalues forming the 9-mode ROM basis are highlighted: 4 real (circles) and 5 complex pairs (squares). The complex eigenvalues correspond to first through fourth bending and first torsion, modified by aerodynamic coupling.}
\label{fig:vfa_eigenbasis}
\end{figure}

The eigenspectrum of the coupled Jacobian at the trimmed equilibrium is shown in~\Cref{fig:vfa_eigenbasis}. The 9 retained eigenvalues are highlighted: 4 real eigenvalues near the origin and 5 complex conjugate pairs. The basis selection follows the strategy outlined in~\Cref{sec:basis}.

The 9 eigenvalues retained in the ROM basis are detailed in~\Cref{tab:rom_eigenvalues}.

\begin{table}[htbp]
\centering
\caption{Reduced-order model eigenvalues (rad/s).}
\label{tab:rom_eigenvalues}
\begin{tabular}{@{}clcc@{}}
\toprule
Number & Mode & Real part & Imaginary part \\
\midrule
1 & gust/fluid/rigid & $-4.147 \times 10^{-2}$ & 0.000 \\
2 & gust/fluid/rigid & $-8.285 \times 10^{-1}$ & 0.000 \\
3 & gust/fluid/rigid & $-2.087$ & 0.000 \\
4 & gust/fluid/rigid & $-2.143$ & 0.000 \\
5 & 1st bending & $-3.715 \times 10^{-1}$ & 1.065 \\
6 & 2nd bending & $-4.695 \times 10^{-1}$ & 4.398 \\
7 & 1st torsion & $-2.918 \times 10^{-1}$ & 5.896 \\
8 & 3rd bending & $-5.817 \times 10^{-1}$ & $1.194 \times 10^{1}$ \\
9 & 4th bending & $-2.383 \times 10^{-1}$ & $1.438 \times 10^{1}$ \\
\bottomrule
\end{tabular}
\end{table}

\subsubsection{ROM convergence study}

\begin{figure}[htbp]
\centering
\includegraphics[width=0.7\textwidth]{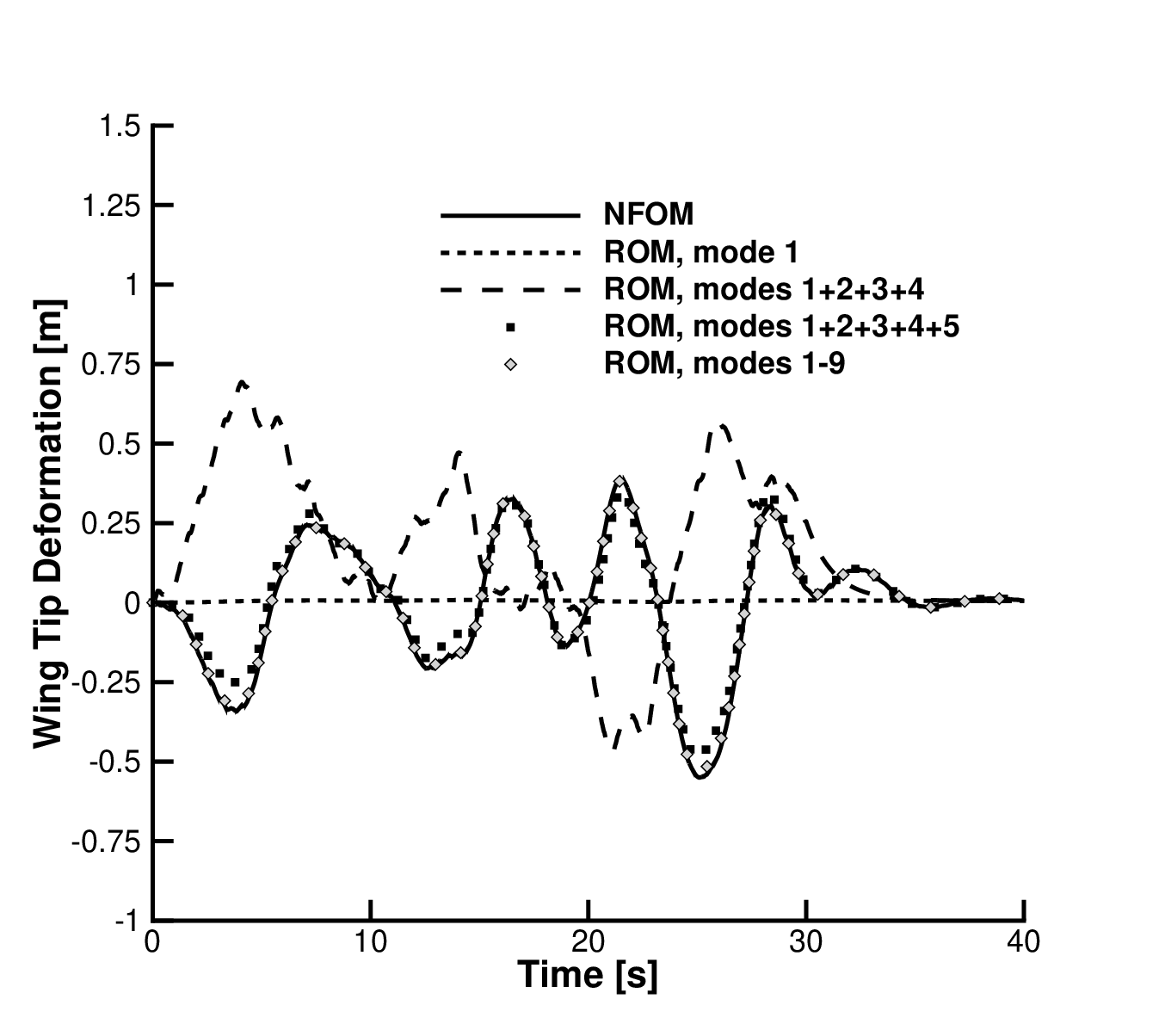}
\caption{Convergence of the ROM wing-tip vertical displacement response with increasing number of retained modes for a Von K\'{a}rm\'{a}n stochastic gust (750~m length-scale). The nonlinear full-order model (NFOM, solid) is progressively approached as modes are added: 1 mode (dotted), 4 modes (long dashed), 5 modes (filled squares), and 9 modes (open diamonds, overlapping with NFOM).}
\label{fig:vfa_convergence}
\end{figure}

The convergence of the ROM with respect to the number of retained modes is demonstrated in~\Cref{fig:vfa_convergence}. Wing-tip vertical displacement response to a stochastic Von K\'{a}rm\'{a}n gust of 750~m length-scale (30~s total simulation) is computed using ROMs with progressively increasing numbers of modes. With a single mode only (mode~1, dotted line near zero), the ROM captures the rigid-body angle-of-attack response but not the structural aeroelastic dynamics. With modes 1--4 (rigid-body and gust coupling, long dashed), the ROM reproduces the slow phugoid-like oscillation but overpredicts the amplitude since structural modes are missing. Including the first bending mode (5 modes, filled squares) recovers the dominant wing-tip oscillation frequency and amplitude. The full 9-mode basis (open diamonds) produces a fully converged solution that overlaps with the nonlinear full-order model (NFOM, solid line).

\subsubsection{Computational speedup}

A parametric study of 37 ``1-minus-cosine'' gust cases is performed, with gust durations ranging from $t_g = 0.5$~s to $t_g = 20.0$~s (corresponding to gust lengths from 12.5~m to 500~m). Each full-order nonlinear simulation requires approximately 6 hours of wall-clock time on a single processor, resulting in a total of 222 hours for the complete parameter sweep. The same 37 cases computed with the 9-mode nonlinear ROM require a total of 22 minutes, achieving a computational speedup of \textbf{600$\times$}.

\begin{figure}[htbp]
\centering
\includegraphics[width=0.6\textwidth]{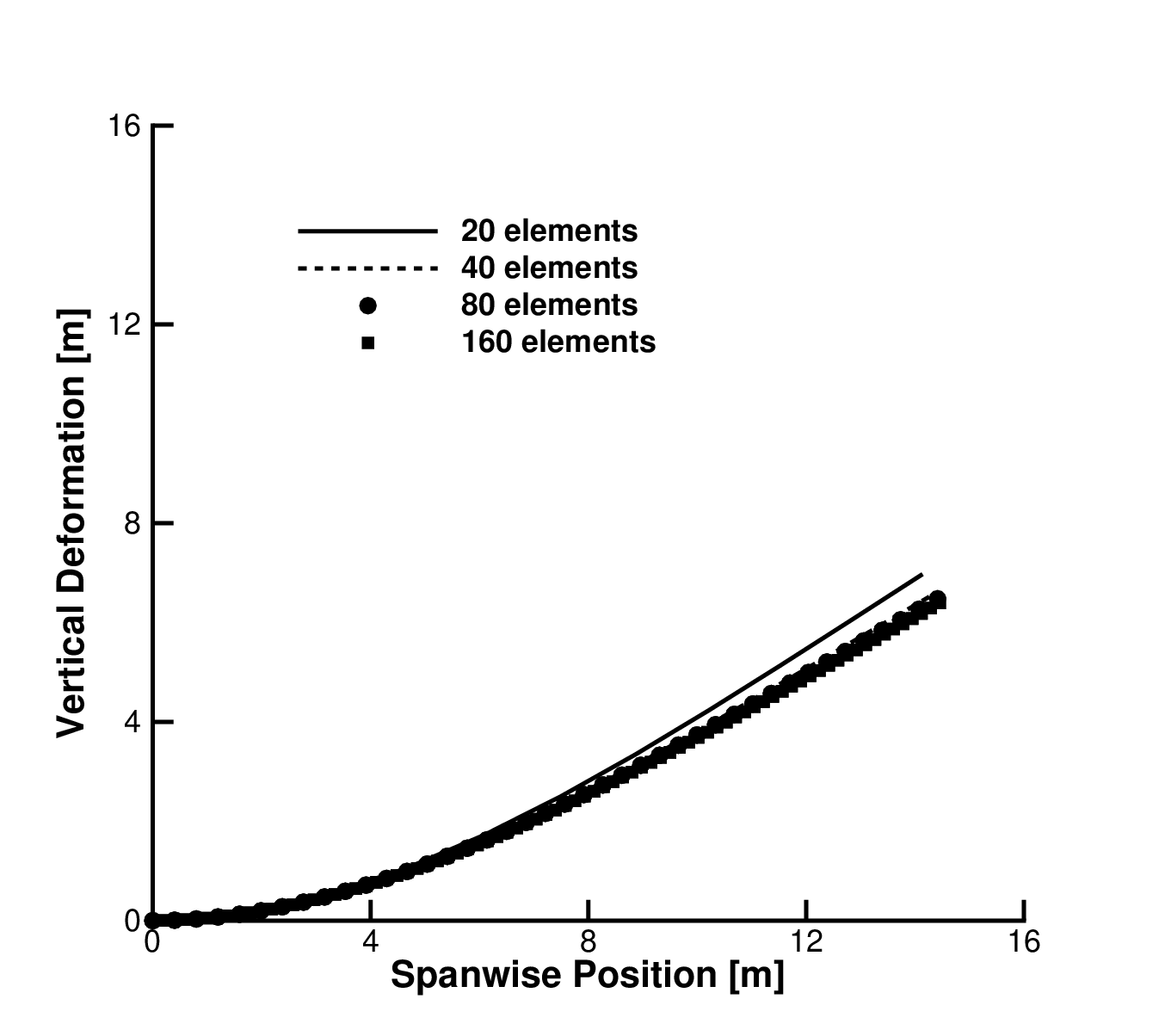}
\caption{Mesh convergence of the static aeroelastic wing deformation for the very flexible flying-wing at $\rho_\infty = 0.25$~kg/m$^3$, $U = 25$~m/s, and $\alpha = 3^\circ$ (clamped condition): 20 elements (solid line), 40 elements (dashed), 80 elements (circles), and 160 elements (squares). The 80-element mesh used for the ROM study provides a well-converged solution.}
\label{fig:vfa_static}
\end{figure}

A mesh convergence study for the static aeroelastic wing deformation is presented in~\Cref{fig:vfa_static}, conducted at $\rho_\infty = 0.25$~kg/m$^3$ (altitude $h = 13{,}500$~m), $U = 25$~m/s, and an initial angle of attack of $3^\circ$ without including the rigid-body degrees of freedom. The results confirm that the selected discretisation of 80 elements provides a converged solution, with negligible difference from the 160-element mesh.

\subsubsection{Linear versus nonlinear ROM accuracy}

\begin{figure}[htbp]
\centering
\begin{subfigure}[b]{0.48\textwidth}
\includegraphics[width=\textwidth]{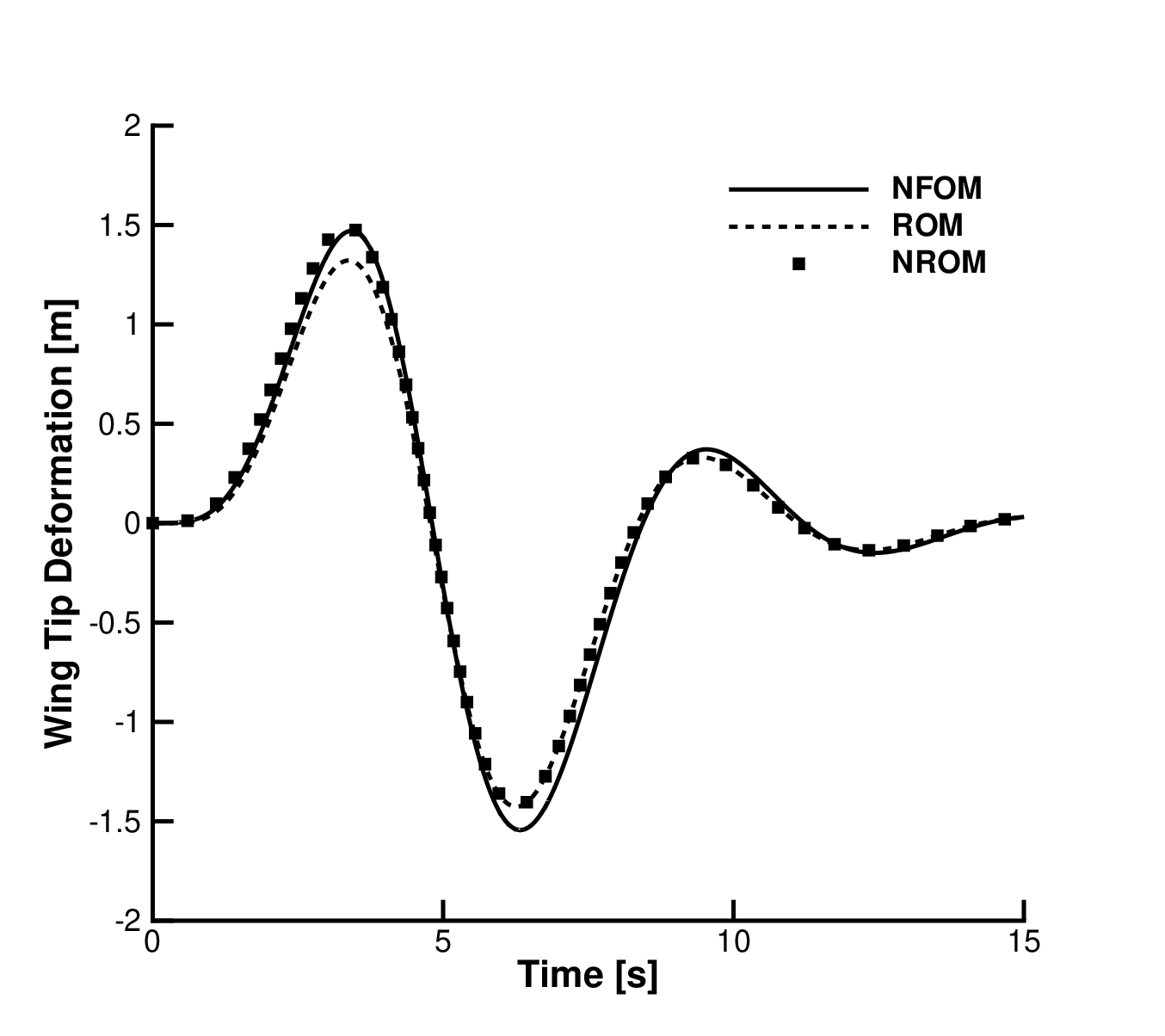}
\caption{Wing-tip vertical displacement}
\label{fig:vfa_wtip_nonlinear}
\end{subfigure}
\hfill
\begin{subfigure}[b]{0.48\textwidth}
\includegraphics[width=\textwidth]{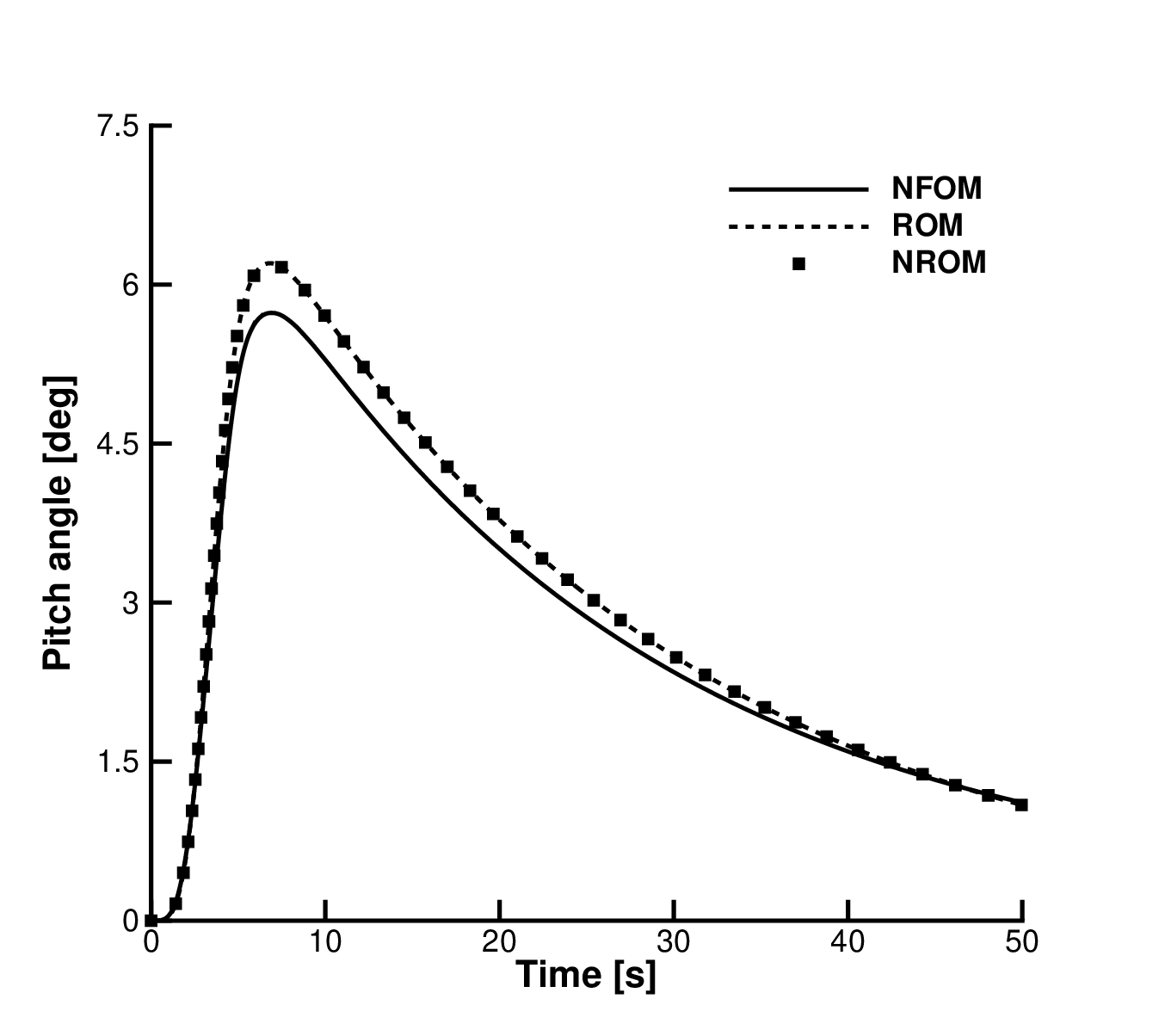}
\caption{Body-frame pitch angle}
\label{fig:vfa_aoa_nonlinear}
\end{subfigure}
\caption{Response of the VFA flying-wing to the worst-case discrete gust (normalised intensity 1.25, $t_g = 5.0$~s): nonlinear full-order model (NFOM, solid), linear ROM (ROM, dashed), and nonlinear ROM with 2nd-order Taylor expansion (NROM, filled squares). The linear ROM slightly overpredicts the peak response amplitude due to the missing geometric stiffening, while the nonlinear ROM closely tracks the full-order solution throughout.}
\label{fig:vfa_nonlinear}
\end{figure}

A critical assessment of ROM accuracy as a function of wing deformation level is presented in~\Cref{fig:vfa_wtip_nonlinear,fig:vfa_aoa_nonlinear}. The linear ROM (retaining only the Jacobian term in the Taylor expansion) accurately predicts the full-order response for wing-tip deformations below approximately 10\% of the wingspan. For larger deformations, the geometric stiffening effect, whereby the wing becomes stiffer as it bends upward due to the change in the lift vector direction, is not captured by the linear approximation, leading to an overestimation of the dynamic response amplitude.

The nonlinear ROM with the second-order Taylor expansion accurately tracks the full-order solution even when the wing tip deformation reaches around 10\% of the wingspan~\citep{Tantaroudas2017bookchapter}. The second-order expansion is sufficient because the quadratic interaction terms $D_{kij} z_i z_j$ in~\Cref{eq:rom} generate effective cubic nonlinearities through the interplay of multiple modal amplitudes, replicating the geometric stiffening behaviour of the exact beam theory.

Third-order Taylor expansion terms ($E_{kijl}$) provide negligible improvement over the second-order model for all gust cases considered. This is a significant computational advantage, as the third-order terms require $\mathcal{O}(m^3) = 729$ additional residual evaluations compared to $\mathcal{O}(m^2) = 81$ for the second-order terms.

\subsubsection{Flexibility effect on flight dynamics}
\label{sec:flex_effect}

The flexibility effect on the coupled flight dynamic response is examined. The coupled system, initially at the equilibrium condition, encounters a strong ``1-minus-cosine'' gust of 125~m length and a maximum velocity of 0.8 of the freestream speed, which can cause large wing deformation and thus affect the flight dynamic response of the vehicle when compared to a rigid flying wing. The varying parameter is the flexibility parameter $\sigma$ which scales the stiffness matrix. Starting from a very stiff wing ($\sigma = 0.001$) and nominal flexibility ($\sigma = 1$), the flexibility parameter is increased up to 4 times.

\begin{figure}[htbp]
\centering
\includegraphics[width=0.45\textwidth]{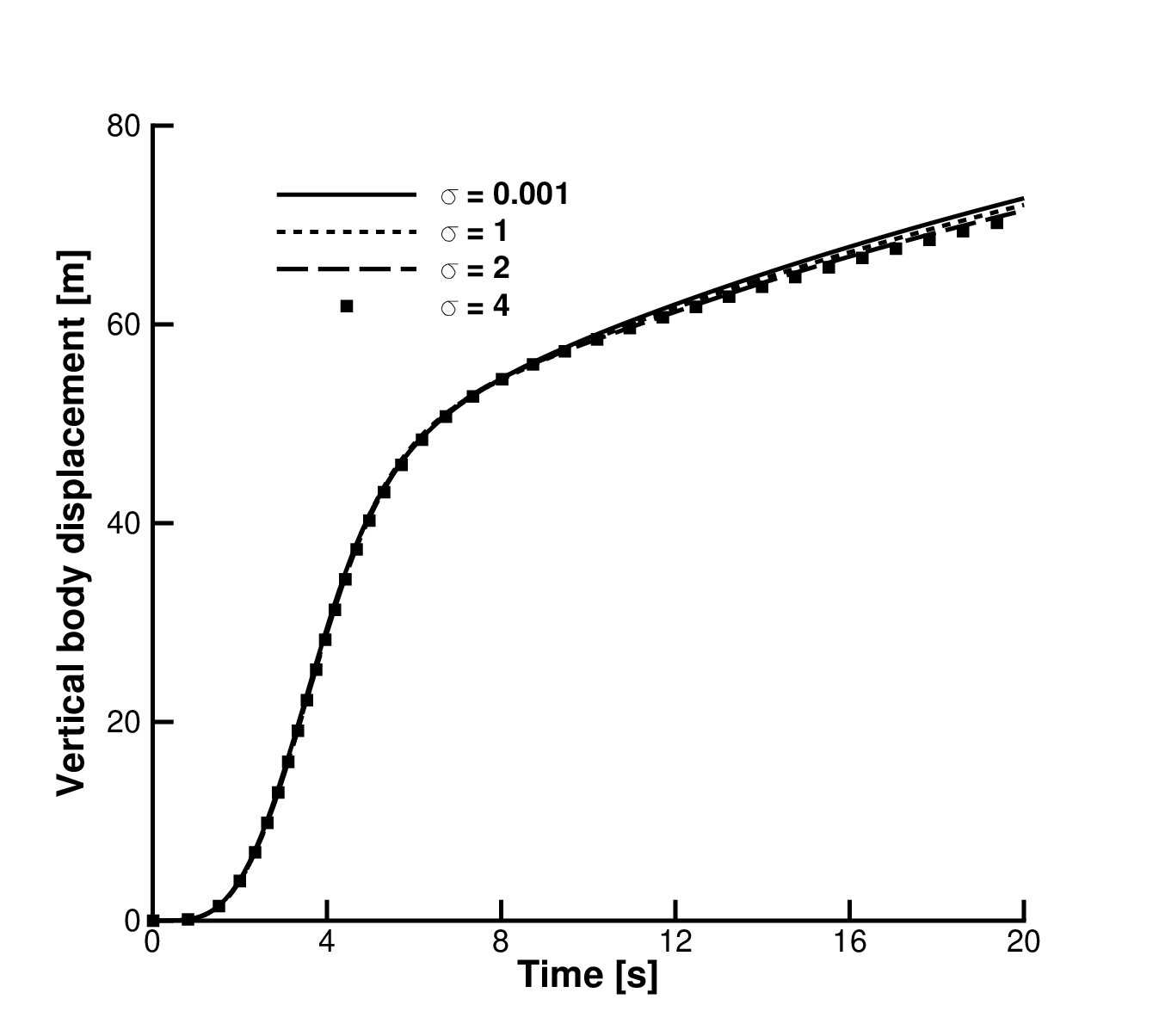}
\includegraphics[width=0.45\textwidth]{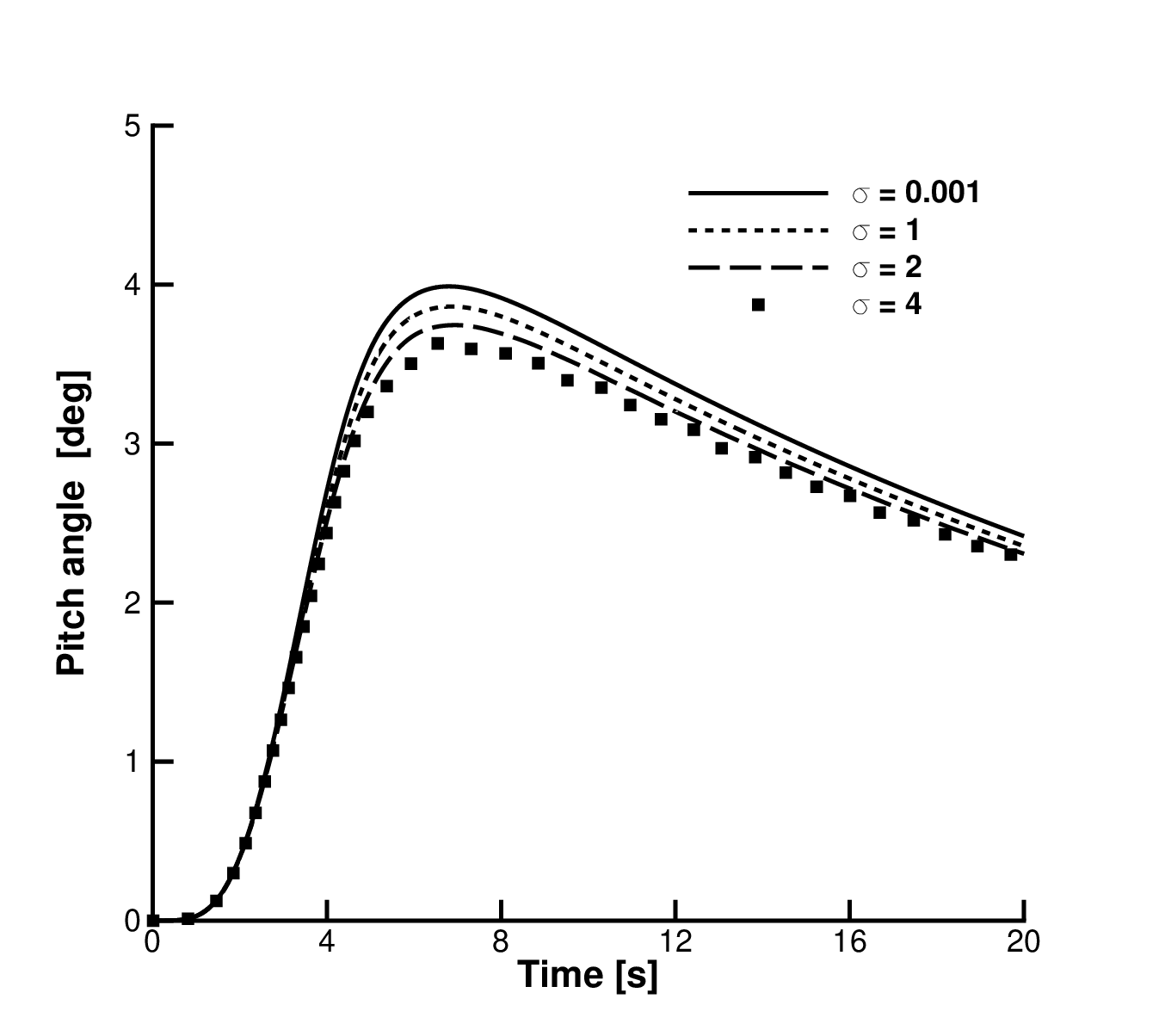}\\
\includegraphics[width=0.45\textwidth]{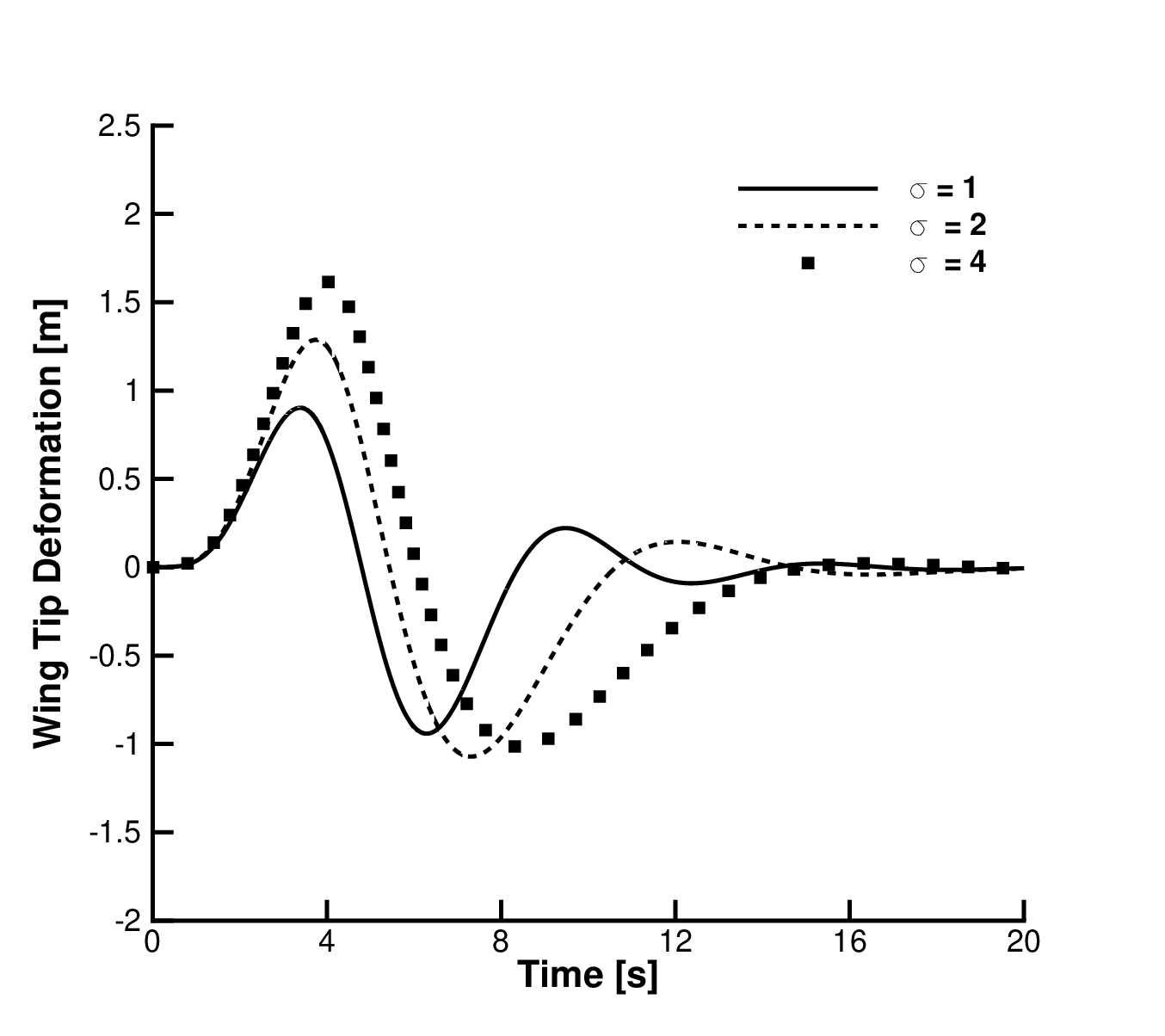}
\includegraphics[width=0.45\textwidth]{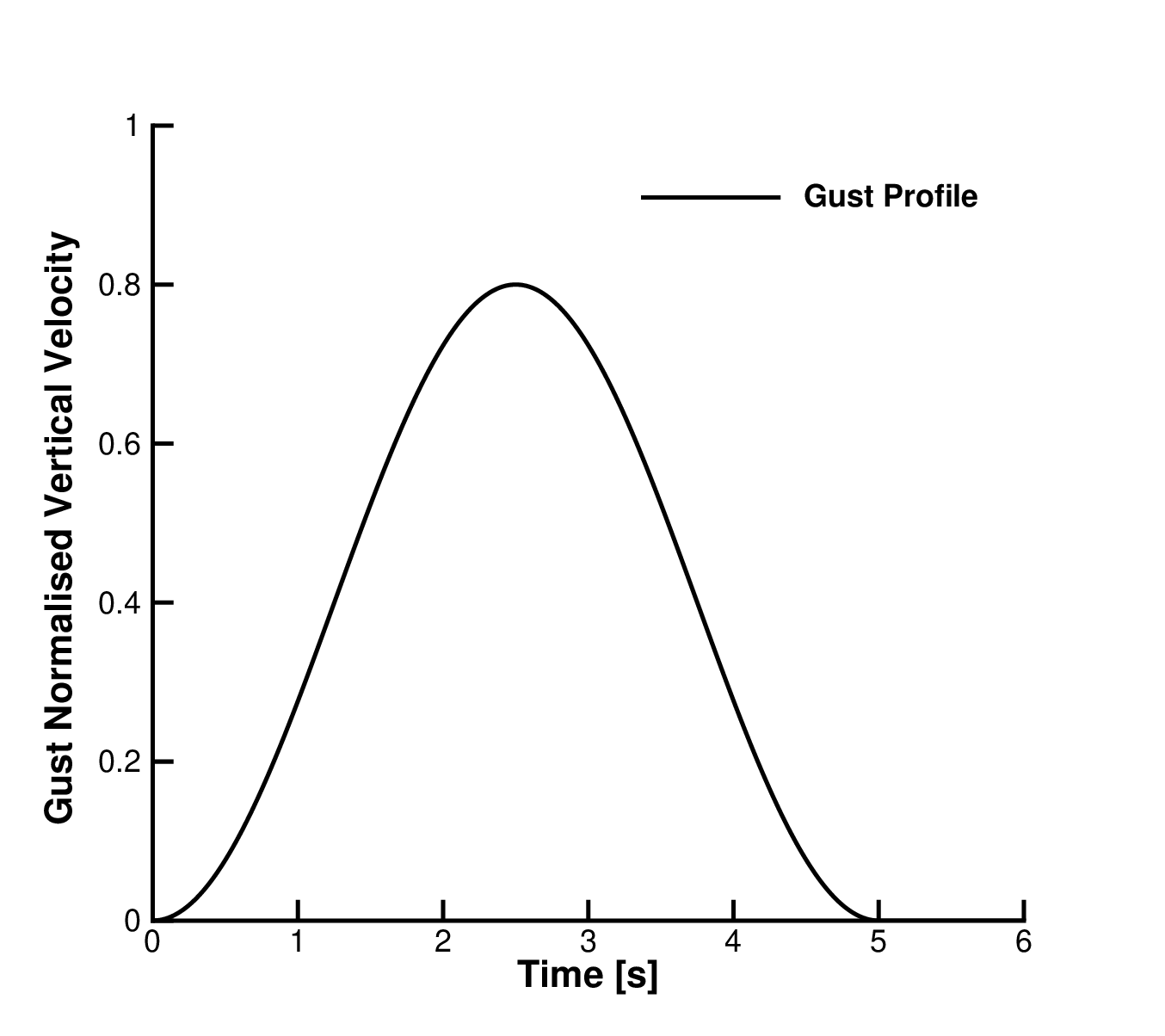}
\caption{Flight dynamics gust response for increasing stiffness parameter $\sigma$ at ($\rho_\infty = 0.25$~kg/m$^3$, $U = 25$~m/s).}
\label{fig:flex_effect}
\end{figure}

The vertical displacement, angle of attack and wing tip deformation are shown in~\Cref{fig:flex_effect} for four different stiffness parameters. When increasing the flexibility, the wing tip deformation increases, affecting the aerodynamic forces and as a result marginally affecting the flight dynamic response. This confirms earlier findings by~\citet{Patil2001}.

\subsubsection{Summary of NMOR performance}

\Cref{tab:nmor_performance} summarises the NMOR performance metrics across all three test cases.

\begin{table}[htbp]
\centering
\caption{Summary of NMOR performance across the three test cases.}
\label{tab:nmor_performance}
\begin{tabular}{@{}lcccc@{}}
\toprule
Test case & FOM dim.\ ($n$) & ROM dim.\ ($m$) & Speedup & Max deformation \\
\midrule
3-DOF aerofoil & 14 & 4 & modest & N/A (2-D) \\
HALE aircraft & 2,016 & 9 & significant & large \\
VFA flying-wing & 1,616 & 9 & 600$\times$ & ${\sim}10\%$ wingspan \\
\bottomrule
\end{tabular}
\end{table}

\section{Discussion}
\label{sec:discussion}

The results demonstrate several important properties of the NMOR framework that merit further discussion.

\textbf{Sufficiency of second-order Taylor expansion.} Across all test cases, retaining only the second-order (quadratic) term in the Taylor expansion provides accuracy comparable to the full nonlinear model, even for cubic structural nonlinearities. This can be understood by noting that the projection of the quadratic operator $\mathcal{B}$ onto the eigenvector basis generates terms of the form $D_{kij} z_i z_j$. When distinct mode pairs interact (e.g., $z_1 z_2$), the effective force on mode $k$ mimics a cubic dependence on the physical displacement amplitude. This mathematical property was first observed in~\citep{DaRonch2013control} and rigorously confirmed in the book chapter~\citep{Tantaroudas2017bookchapter}.

\textbf{Formulation independence.} The NMOR approach requires only: (i) evaluation of the residual $\mathbf{R}(\mathbf{w})$ at arbitrary states, and (ii) solution of the eigenvalue problem for the Jacobian $\mathbf{A}$. It does not require access to the internal structure of the full-order model, making it applicable to CFD-coupled systems~\citep{DaRonch2014scitech_flight}, UVLM-based models~\citep{Murua2012}, or commercial FE codes.

\textbf{Basis selection strategy.} The two-tier selection (real eigenvalues for rigid-body/gust modes, lightly damped complex eigenvalues for structural modes) is physically motivated and has proven robust across all configurations tested. The convergence study in~\Cref{fig:vfa_convergence} confirms that 9 modes are sufficient for the flying-wing configuration, with the first bending mode being the most critical addition beyond the rigid-body dynamics.

\textbf{Limitations.} The current framework assumes a single equilibrium point for the Taylor expansion. For flight regimes involving large changes in trim state (e.g., aggressive manoeuvres), the ROM may require updating. Multi-point expansion strategies, employing Taylor expansions about multiple equilibria, offer a potential remedy. Additionally, the strip-theory aerodynamic model limits accuracy at high angles of attack or for configurations with strong three-dimensional flow effects, as evidenced by the HALE validation (\Cref{fig:hale_static}).

\section{Conclusions}
\label{sec:conclusions}

A systematic nonlinear model order reduction framework for coupled aeroelastic-flight dynamic systems has been presented and validated on three test cases of increasing complexity. The key findings are:

\begin{enumerate}
\item The NMOR technique achieves reductions from $\sim$1,600 degrees of freedom to 9, with computational speedups of up to 600$\times$ for parametric gust studies (222 hours reduced to 22 minutes).
\item The methodology is independent of the original full-order model formulation, it requires only a coupled system expressed in first-order state-space form and the ability to evaluate residuals and Jacobian eigenvalues.
\item The second-order Taylor expansion is sufficient for capturing cubic structural nonlinearities arising from geometrically-exact beam theory, eliminating the need for computationally expensive third-order terms. This is because quadratic modal interactions effectively reproduce the cubic stiffening behaviour.
\item A systematic eigenvector basis selection strategy is demonstrated: real eigenvalues near the origin for rigid-body and gust coupling, followed by lightly damped complex eigenvalues for structural modes. Convergence is achieved with 9 modes for the most complex configuration tested.
\item The framework accurately captures nonlinear phenomena including limit-cycle oscillations (3-DOF aerofoil) and large wing deformations exceeding 10\% of the wingspan (VFA flying-wing) that are beyond the validity of linear methods.
\item The aeroelastic framework was validated for a rigid and a flexible aircraft against published data. The trimming in the vertical equilibrium differed greatly between rigid and flexible configurations, and these differences grew larger as the flexibility increased, consistent with past studies.
\end{enumerate}

The reduced-order models generated by this approach enable rapid parametric studies for very flexible aircraft, including gust response predictions across a wide range of flight conditions. The reduced-order models allow simulating considerably longer time durations than those generally considered in routine analyses, enabling efficient verification of aeroelastic system stability. The framework provides a practical computational tool for the design and certification of next-generation HALE and solar-powered platforms, bridging the gap between high-fidelity nonlinear analysis and the need for rapid, repeated evaluations in the design loop.

\section*{Acknowledgements}

This work was supported by the U.K.\ Engineering and Physical Sciences Research Council (EPSRC) grant EP/I014594/1 on ``Nonlinear Flexibility Effects on Flight Dynamics and Control of Next-Generation Aircraft.'' The authors are grateful to Prof.\ A.\ Da Ronch and Prof.\ K.J.\ Badcock for their valuable guidance on flight dynamics modelling.

\bibliographystyle{unsrtnat}
\bibliography{references}

@INBOOK{Tantaroudas2017bookchapter,
  author    = {Tantaroudas, N.D. and Da Ronch, A.},
  title     = {Nonlinear Reduced-order Aeroservoelastic Analysis of Very Flexible Aircraft},
  booktitle = {Advanced UAV Aerodynamics, Flight Stability and Control},
  publisher = {John Wiley \& Sons, Ltd},
  address   = {Chichester, UK},
  chapter   = {4},
  pages     = {143--179},
  year      = {2017},
  doi       = {10.1002/9781118928691.ch4}
}

@INPROCEEDINGS{Tantaroudas2015scitech,
  author    = {Tantaroudas, N.D. and Da Ronch, A. and Badcock, K.J. and Palacios, R.},
  title     = {Model Order Reduction for Control Design of Flexible Free-Flying Aircraft},
  booktitle = {AIAA Atmospheric Flight Mechanics Conference, AIAA SciTech 2015},
  series    = {AIAA Paper 2015-0240},
  year      = {2015},
  doi       = {10.2514/6.2015-0240}
}

@INPROCEEDINGS{Tantaroudas2014aviation,
  author    = {Tantaroudas, N.D. and Da Ronch, A. and Gai, G. and Badcock, K.J. and Palacios, R.},
  title     = {An Adaptive Aeroelastic Control Approach using Non Linear Reduced Order Models},
  booktitle = {14th AIAA Aviation Technology, Integration, and Operations Conference},
  series    = {AIAA Paper 2014-2590},
  year      = {2014},
  doi       = {10.2514/6.2014-2590}
}

@INPROCEEDINGS{DaRonch2014scitech_flight,
  author    = {Da Ronch, A. and McCracken, A.J. and Tantaroudas, N.D. and Badcock, K.J. and Hesse, H. and Palacios, R.},
  title     = {Assessing the Impact of Aerodynamic Modelling on Manoeuvring Aircraft},
  booktitle = {AIAA SciTech 2014, AIAA Atmospheric Flight Mechanics Conference},
  series    = {AIAA Paper 2014-0732},
  year      = {2014},
  doi       = {10.2514/6.2014-0732}
}

@INPROCEEDINGS{DaRonch2014flutter,
  author    = {Da Ronch, A. and Tantaroudas, N.D. and Jiffri, S. and Mottershead, J.E.},
  title     = {A Nonlinear Controller for Flutter Suppression: from Simulation to Wind Tunnel Testing},
  booktitle = {55th AIAA/ASME/ASCE/AHS/SC Structures, Structural Dynamics, and Materials Conference},
  series    = {AIAA Paper 2014-0345},
  year      = {2014},
  doi       = {10.2514/6.2014-0345}
}

@INPROCEEDINGS{Fichera2014isma,
  author    = {Fichera, S. and Jiffri, S. and Wei, X. and Da Ronch, A. and Tantaroudas, N.D. and Mottershead, J.E.},
  title     = {Experimental and Numerical Study of Nonlinear Dynamic Behaviour of an Aerofoil},
  booktitle = {ISMA 2014 Conference on Noise and Vibration Engineering},
  pages     = {3609--3618},
  year      = {2014}
}

@INPROCEEDINGS{DaRonch2013gust,
  author    = {Da Ronch, A. and Tantaroudas, N.D. and Timme, S. and Badcock, K.J.},
  title     = {Model Reduction for Linear and Nonlinear Gust Loads Analysis},
  booktitle = {54th AIAA/ASME/ASCE/AHS/ASC Structures, Structural Dynamics, and Materials Conference},
  series    = {AIAA Paper 2013-1492},
  year      = {2013},
  doi       = {10.2514/6.2013-1492}
}

@INPROCEEDINGS{DaRonch2013control,
  author    = {Da Ronch, A. and Tantaroudas, N.D. and Badcock, K.J.},
  title     = {Reduction of Nonlinear Models for Control Applications},
  booktitle = {54th AIAA/ASME/ASCE/AHS/ASC Structures, Structural Dynamics, and Materials Conference},
  series    = {AIAA Paper 2013-1491},
  year      = {2013},
  doi       = {10.2514/6.2013-1491}
}

@INPROCEEDINGS{Papatheou2013ifasd,
  author    = {Papatheou, E. and Tantaroudas, N.D. and Da Ronch, A. and Cooper, J.E. and Mottershead, J.E.},
  title     = {Active Control for Flutter Suppression: an Experimental Investigation},
  booktitle = {International Forum on Aeroelasticity and Structural Dynamics},
  series    = {IFASD Paper 2013-8D},
  year      = {2013}
}

@ARTICLE{Patil2001,
  author  = {Patil, M.J. and Hodges, D.H. and Cesnik, C.E.S.},
  title   = {Nonlinear Aeroelasticity and Flight Dynamics of High-Altitude Long-Endurance Aircraft},
  journal = {Journal of Aircraft},
  volume  = {38},
  number  = {1},
  pages   = {88--94},
  year    = {2001},
  doi     = {10.2514/2.2738}
}

@ARTICLE{Patil2006,
  author  = {Patil, M.J. and Hodges, D.H.},
  title   = {Flight Dynamics of Highly Flexible Flying Wings},
  journal = {Journal of Aircraft},
  volume  = {43},
  number  = {6},
  pages   = {1790--1799},
  year    = {2006},
  doi     = {10.2514/1.17640}
}

@ARTICLE{Murua2012,
  author  = {Murua, J. and Palacios, R. and Graham, J.M.R.},
  title   = {Applications of the Unsteady Vortex-Lattice Method in Aircraft Aeroelasticity and Flight Dynamics},
  journal = {Progress in Aerospace Sciences},
  volume  = {55},
  pages   = {46--72},
  year    = {2012},
  doi     = {10.1016/j.paerosci.2012.06.001}
}

@ARTICLE{Hesse2014,
  author  = {Hesse, H. and Palacios, R.},
  title   = {Reduced-Order Aeroelastic Models for Dynamics of Maneuvering Flexible Aircraft},
  journal = {AIAA Journal},
  volume  = {52},
  number  = {8},
  pages   = {1717--1732},
  year    = {2014},
  doi     = {10.2514/1.J052684}
}

@INPROCEEDINGS{DaRonch2012rom,
  author  = {Da Ronch, A. and Badcock, K.J. and Wang, Y. and Wynn, A. and Palacios, R.},
  title   = {Nonlinear Model Reduction for Flexible Aircraft Control Design},
  booktitle = {AIAA Atmospheric Flight Mechanics Conference},
  series  = {AIAA Paper 2012-4404},
  year    = {2012},
  doi     = {10.2514/6.2012-4404}
}

@ARTICLE{Theodorsen1935,
  author  = {Theodorsen, T.},
  title   = {General Theory of Aerodynamic Instability and the Mechanism of Flutter},
  journal = {NACA Report},
  volume  = {496},
  year    = {1935}
}

@ARTICLE{Wagner1925,
  author  = {Wagner, H.},
  title   = {{\"U}ber die Entstehung des dynamischen Auftriebes von Tragfl{\"u}geln},
  journal = {Zeitschrift f{\"u}r Angewandte Mathematik und Mechanik},
  volume  = {5},
  number  = {1},
  pages   = {17--35},
  year    = {1925}
}

@ARTICLE{Kussner1936,
  author  = {K{\"u}ssner, H.G.},
  title   = {Zusammenfassender Bericht {\"u}ber den instation{\"a}ren Auftrieb von Fl{\"u}geln},
  journal = {Luftfahrtforschung},
  volume  = {13},
  pages   = {410--424},
  year    = {1936}
}

@ARTICLE{Dowell2001,
  author  = {Dowell, E.H. and Hall, K.C.},
  title   = {Modeling of Fluid-Structure Interaction},
  journal = {Annual Review of Fluid Mechanics},
  volume  = {33},
  pages   = {445--490},
  year    = {2001},
  doi     = {10.1146/annurev.fluid.33.1.445}
}

@ARTICLE{Lucia2004,
  author  = {Lucia, D.J. and Beran, P.S. and Silva, W.A.},
  title   = {Reduced-Order Modeling: New Approaches for Computational Physics},
  journal = {Progress in Aerospace Sciences},
  volume  = {40},
  number  = {1-2},
  pages   = {51--117},
  year    = {2004},
  doi     = {10.1016/j.paerosci.2003.12.001}
}

@ARTICLE{Palacios2010,
  author  = {Palacios, R.},
  title   = {Nonlinear Normal Modes in an Intrinsic Theory of Anisotropic Beams},
  journal = {Journal of Sound and Vibration},
  volume  = {330},
  number  = {8},
  pages   = {1772--1792},
  year    = {2011},
  doi     = {10.1016/j.jsv.2010.10.023}
}

@ARTICLE{Hodges2003,
  author  = {Hodges, D.H.},
  title   = {Geometrically Exact, Intrinsic Theory for Dynamics of Curved and Twisted Anisotropic Beams},
  journal = {AIAA Journal},
  volume  = {41},
  number  = {6},
  pages   = {1131--1137},
  year    = {2003},
  doi     = {10.2514/2.2054}
}

@ARTICLE{Badcock2011,
  author  = {Badcock, K.J. and Timme, S. and Marques, S. and Khodaparast, H. and Palacios, R. and Mughal, M.S. and Woodgate, M.A.},
  title   = {Transonic Aeroelastic Simulation for Instability Searches and Uncertainty Analysis},
  journal = {Progress in Aerospace Sciences},
  volume  = {47},
  number  = {5},
  pages   = {392--423},
  year    = {2011},
  doi     = {10.1016/j.paerosci.2011.05.002}
}

@ARTICLE{Jones1938,
  author  = {Jones, R.T.},
  title   = {Operational Treatment of the Non-Uniform Lift Theory in Airplane Dynamics},
  journal = {NACA Technical Note},
  volume  = {667},
  year    = {1938}
}

@ARTICLE{Noll2004,
  author  = {Noll, T.E. and Brown, J.M. and Perez-Davis, M.E. and Ishmael, S.D. and Tiffany, G.C. and Gaier, M.},
  title   = {Investigation of the {Helios} Prototype Aircraft Mishap},
  journal = {NASA Report},
  year    = {2004}
}

@ARTICLE{SuCesnik2010,
  author  = {Su, W. and Cesnik, C.E.S.},
  title   = {Nonlinear Aeroelasticity of a Very Flexible Blended-Wing-Body Aircraft},
  journal = {Journal of Aircraft},
  volume  = {47},
  number  = {5},
  pages   = {1539--1553},
  year    = {2010},
  doi     = {10.2514/1.47317}
}

@ARTICLE{Alighanbari1995,
  author  = {Alighanbari, H. and Price, S.J.},
  title   = {The post-Hopf-bifurcation response of an airfoil in incompressible two-dimensional flow},
  journal = {Nonlinear Dynamics},
  volume  = {10},
  pages   = {381--400},
  year    = {1996}
}

@ARTICLE{Irani2011,
  author  = {Irani, S. and Sarrafzadeh, H. and Amoozgar, M.R.},
  title   = {Bifurcation in a 3-DOF Airfoil with Cubic Structural Nonlinearity},
  journal = {Chinese Journal of Aeronautics},
  volume  = {24},
  pages   = {265--278},
  year    = {2011}
}

@ARTICLE{Riso2023joa,
  author  = {Riso, C. and Cesnik, C.E.S.},
  title   = {Impact of Low-Order Modeling on Aeroelastic Predictions for Very Flexible Wings},
  journal = {Journal of Aircraft},
  volume  = {60},
  number  = {3},
  pages   = {662--687},
  year    = {2023},
  doi     = {10.2514/1.C036869}
}

@ARTICLE{Riso2023jfs,
  author  = {Riso, C. and Cesnik, C.E.S.},
  title   = {Geometrically Nonlinear Effects in Wing Aeroelastic Dynamics at Large Deflections},
  journal = {Journal of Fluids and Structures},
  volume  = {120},
  pages   = {103897},
  year    = {2023},
  doi     = {10.1016/j.jfluidstructs.2023.103897}
}

@ARTICLE{Candon2024,
  author  = {Candon, M. and Hale, E. and Balajewicz, M. and Delgado-Gutierrez, A. and Marzocca, P.},
  title   = {Parameterization of Nonlinear Aeroelastic Reduced Order Models via Direct Interpolation of {Taylor} Partial Derivatives},
  journal = {Nonlinear Dynamics},
  volume  = {112},
  pages   = {17649--17670},
  year    = {2024},
  doi     = {10.1007/s11071-024-09976-z}
}

@ARTICLE{Goizueta2022,
  author  = {Goizueta, N. and Wynn, A. and Palacios, R.},
  title   = {Adaptive Sampling for Interpolation of Reduced-Order Aeroelastic Systems},
  journal = {AIAA Journal},
  volume  = {60},
  number  = {11},
  pages   = {6183--6202},
  year    = {2022},
  doi     = {10.2514/1.J062050}
}

\end{document}